\documentclass[12pt,epsf]{report}
\input  epsf.tex
\usepackage{graphicx}
\catcode`@=11     
     
\newdimen\jot \jot=3pt
\newskip\z@skip \z@skip=0pt plus0pt minus0pt
\newdimen\z@ \z@=0pt 
\dimendef\dimen@=0
     
\def\m@th{\mathsurround=\z@}
     
\def\ialign{\everycr{}\tabskip\z@skip\halign} 
     
\def\openup{\afterassignment\@penup\dimen@=}
\def\@penup{\advance\lineskip\dimen@
  \advance\baselineskip\dimen@
  \advance\lineskiplimit\dimen@}

\def\mathhyphen{{\the\textfont0 \char`\-}}

\def\eqalign#1{\null\,\vcenter{\openup\jot\m@th
  \ialign{\strut\hfil$\displaystyle{##}$&$\displaystyle{{}##}$\hfil
      \crcr#1\crcr}}\,}

\def\pc{\varphi_c}
\def\vev{\langle \phi \rangle}

\begin{document}

\baselineskip 18pt

\centerline{RADIATIVE CORRECTIONS AS THE ORIGIN OF}
\centerline{SPONTANEOUS SYMMETRY BREAKING}

\bigskip

\centerline{A thesis presented}
\medskip

\centerline{by}
\medskip

\centerline{Erick James Weinberg}
\medskip

\centerline{to}

\centerline{The Department of Physics}

\centerline{in partial fulfillment of the requirements}
\centerline{for the degree of}
\centerline{Doctor of Philosophy}
\centerline{in the subject of}
\centerline{Physics}

\bigskip

\centerline{Harvard University}
\centerline{Cambridge, Massachusetts}
\centerline{May 1973}

\setcounter{page}{0}
\thispagestyle{empty}

\eject

\centerline{Abstract}

\bigskip

Using a functional formalism, we investigate the effect of radiative
corrections on the possibility of spontaneous symmetry breaking.  We
find that in some models, in particular, massless gauge theories with
scalar mesons, the radiative corrections can induce spontaneous
symmetry breaking, even though the classical approximation indicates
that the vacuum is symmetric.  Among the consequences of this
phenomenon is a relationship between the masses of the scalar and
vector mesons, predicting (for small coupling constants) that the
scalar mesons are much lighter.  We also apply our analysis to models
in which the classical approximation indicates an asymmetric vacuum,
including one in which our methods are particularly useful because the
classical approximation does not completely specify the nature of the
vacuum.  It is possible to improve our analysis by the use of 
renormalization group methods; we do this for several models.

\thispagestyle{empty}

\vfill 

\centerline{ii}

\eject

\centerline{Acknowledgments}

\bigskip

I would like to thank my thesis advisor, Professor Sidney Coleman, for
his assistance and encouragement.  I am grateful to him for showing 
me that the best way to solve a difficult problem is often to not do
the obvious.  It has been a privilege to have worked with him.

David Politzer has been of immeasurable assistance in this work; I have 
had innumerable enjoyable and fruitful conversations with him.  In 
addition, he has helped me learn that two people working together 
can calculate Feynman diagrams with neither can calculate alone.

I have profitted greatly from discussions with many faculty members,
especially Professors Thomas Appelquist, Sheldon Glashow, and Helen
Quinn, as well as with Dr. Howard Georgi.

I have learned a great deal from my fellow graduate students.  I 
especially wish to thank Jim Carazzone and Terry Goldman.  

Teaching has been an enjoyable and rewarding experience;
I thank Professors Paul Bamberg and Arthur Jaffe for the opportunity
to have taught with them.

I would like to thank Alan Candiotti for the use of his electric
typewriter; without it, this thesis would never have been finished.

Finally, I wish to thank my wife, Carolyn.  Her love and understanding
have helped me through many of the dark and discouraging moments of 
graduate study.

\thispagestyle{empty}

\vfill \centerline{iii}
\eject

\centerline{Table of Contents}

\bigskip

{\parindent =0pt
Abstract  \hfill  ii

Acknowledgments \hfill  iii 

Chapter I. Introduction  \hfill 1

Chapter II. Formalism \hfill 3

Chapter III. A Simple Example \hfill 10

Chapter IV. Bigger and Better Models \hfill 20

Chapter V. Massive Theories \hfill 33

Chapter VI. The Renormalization Group \hfill 46

Chapter VII. Conclusions \hfill 61

Appendix \hfill 62

Footnotes \hfill 63

Figures \hfill 65}

\thispagestyle{empty}
\setcounter{page}{0}

\vfill \centerline{iv} \eject

\centerline{I. Introduction}

\bigskip

Spontaneous symmetry breaking has been of great importance in the
study of elementary particles during the last decade, first because it
explained why pions were almost massless$^1$, and then because it
explained why weak intermediate vector mesons did not have to be
massless$^2$.  In all of its applications to particle physics,
spontaneous symmetry breaking has been induced by writing the
symmetric Lagrangian in a somewhat unnatural manner --- with a
negative mass-squared term.  In this paper we will show that it is
possible for a theory to have an asymmetric vacuum even when the
Lagrangian does not contain any such unnatural terms; instead, the
spontaneous symmetry breaking is induced by the quantum corrections to
the classical field theory.$^3$ In particular, theories which do not
contain any mass terms may exhibit this phenomenon; one may consider
that the theory takes this form in order to avoid the violent infrared
divergences which would otherwise occur.

There are some interesting consequences when theories of this type
display spontaneous symmetry breaking.  The theory, originally
formulated in terms of dimensionless parameters, can be rewritten to
show explicit dependence on a dimensional quantity, namely the vacuum
expectation value of one of the scalar fields.  One of the
dimensionless parameters can be eliminated, and one obtains a
relationship among other quantities in the theory.  In a gauge theory,
this is generally a relationship among masses, predicting a scalar
meson mass which is much less that the vector meson masses.  We begin
our discussion in Chapter II by describing a formalism which is
particularly suited to our purposes.  In Chapter III we apply our
methods to the study of a particularly simple model, where we do not
find evidence of spontaneous symmetry breaking.  In Chapter IV we
introduce more complicated theories, including some where the vacuum
is asymmetric.  All of the models considered in Chapters III and IV
are without mass terms.  In Chapter V we show how these can be
embedded in the more general class of theories with mass terms (of
either sign).  We also give an example where our methods are
particularly useful, even though the classical treatment of the theory
already predicts spontaneous symmetry breaking.  In Chapter VI we show
how the methods of the renormalization group allow us to improve the
analysis of Chapters III and IV.  We list our conclusions in Chapter
VII.  An appendix discusses the relationship between our results and a
theorem of Georgi and Glashow.

\vfill \eject

\centerline{II. Formalism}

\bigskip

In the usual treatment of spontaneous symmetry breaking in quantum
field theory, one assumes that by looking at the Lagrangian of a
particular theory it is possible to determine whether or not the
theory displays spontaneous symmetry breaking.  That is, one defines
the negative of the non-derivative terms in the Lagrangian to be a
potential, assumes that all fields are constant in space-time in the
vacuum state, and minimizes the potential as a function of the fields
to determine the expectation value of the fields in the vacuum state.
If the vacuum expectation value of some field is non-invariant under
some symmetry of the Lagrangian, spontaneous symmetry breaking is said
to occur.

In a classical field theory, this would be precisely the thing to do.  In
a quantum field theory, this procedure is approximate at best, since the
quantized fields are non-commuting operators and one must be careful in 
manipulating them.  To put things differently, the operator nature of the
fields gives rise to internal loops in Feynman diagrams; that is, there are
both radiative corrections to the interactions in the Lagrangian and new
interactions which are not explicit in the Lagrangian.  These effects are
clearly relevant in determining the nature of the vacuum; the usual 
assumption is that their effect is only to make small quantitative
changes and that the classical approximation gives a correct qualitative
description of the vacuum.  Since we are interested in considering cases
where this assumption may be false, it is desirable to use a formalism 
which treats the classical approximation and the quantum corrections on 
the same basis.

Such a formalism is provided by the functional methods introduced by
Schwinger and developed by Jona-Lasinio, whose treatment we
follow$^4$.  These method enable us to define an effective potential
which includes all the interactions of the theory, both those explicit
in the Lagrangian and those arising from quantum corrections.  
Furthermore, this effective potential is a function of classical fields, 
so it can be easily manipulated.  Finally, its minima determine the 
nature of the ground state of the theory.

One begins by adding to the Lagrangian a coupling to external 
c-number sources, one source for each field in the theory; that is,
$$
  {\cal L}(\phi_i, \partial^\mu \phi_i) \rightarrow 
 {\cal L}(\phi_i, \partial^\mu \phi_i) + \sum_i \phi_i(x) J_i(x) \, ,
                                      \eqno(2.1)
$$ 
where the $\phi_i$ represent all the fields of the theory, whatever their
spin.  Next one defines a functional $W(J)$ in terms of the probability 
amplitude for the vacuum state in the far past to go into the vacuum state
in the far future in the presence of the external sources:
$$  
   e^{iW(J)}  = \langle 0^+ | 0^- \rangle \, .       \eqno(2.2)
$$
The importance of $W(J)$ lies in the fact that it is the generating
functional for the connected Green's functions; that is, we can write
$$\eqalign{
   W(J) =&  \sum_{n_1,n_2, \dots, n_k} 
    {1 \over n_1! n_2! \dots n_k!} 
  \int d^4x_1 d^4x_2 \dots d^4x_{n_1}  \dots d^4w_{n_k} 
    \cr & \, \,  \times 
    G^{(n_1,n_2, \dots, n_k)}(x_1, x_2,\dots, w_{n_k})
    J_1(x_1) J_1(x_2) \dots J_1(x_{n_1}) \dots J_k(w_{n_k})  \, .
   }\eqno(2.3)
$$
Here $G^{(n_1,n_2, \dots, n_k)}(x_1, x_2,\dots, w_{n_k})$ is the sum of 
all connected Feynman diagrams with $n_1$ external lines of type 1, 
$n_2$ external lines of type 2, etc.

Now one defines classical fields $\Phi_{ic}(x)$ by 
$$
   \Phi_{ic}(x) = {\delta W(J) \over \delta J_i(x)} 
     = {\langle 0^+ |\phi_i(x)| 0^- \rangle_J 
        \over \langle 0^+ | 0^- \rangle_J}   \, .
    \eqno(2.4)
$$
Finally, we define the effective action, $\Gamma(\Phi_c)$, by a 
functional Legendre transformation:
$$
  \Gamma(\Phi_c)  = W(J) - \sum_i \int d^4 x J_i(x) \Phi_{ic}(x)
     \, .    \eqno(2.5)
$$
From Eq.~(2.5) we can immediately conclude that 
$$
     {\delta \Gamma(\Phi_c)\over \delta \Phi_{ic}(x)} 
      = - J_i(x)  \, .
     \eqno(2.6)
$$ 
The effective action is also a generating functional; it can be 
written
$$\eqalign{
   \Gamma(\Phi_c) &= \sum_{n_1,n_2, \dots, n_k}
    {1 \over n_1! n_2! \dots n_k!}
  \int d^4x_1 d^4x_2 \dots d^4x_{n_1}  \dots d^4w_{n_k}
    \cr &\, \,  \times
    \Gamma^{(n_1,n_2, \dots, n_k)}(x_1, x_2,\dots, w_{n_k})
    \Phi_{1c}(x_1) \Phi_{1c}(x_2) \dots \Phi_{1c}(x_{n_1}) 
    \dots \Phi_{kc}(w_{n_k})  \, .
   }\eqno(2.7)
$$ 
The $\Gamma^{(n_1,n_2, \dots, n_k)}(x_1, x_2,\dots, w_{n_k})$ are the
IPI (one-particle-irreducible) Green's functions, defined as the sum
of all connected Feynman diagrams which cannot be disconnected by
cutting a single internal line; these are evaluated without
propagators on the external lines.
 
Although Eq.~(2.7) is the usual expansion of the effective action, it is 
not the one best suited for studying the possibility of spontaneous
symmetry breaking.  A better expansion is obtained by expanding in powers
of the external momenta about the point where all external momenta are
zero; in other words, a Taylor series expansion about a constant 
value, $\varphi_{ic}$, of the classical fields $\Phi_{ic}(x)$.  In this
expansion we do not distinguish terms with different numbers of external
scalar particles, but we do distinguish terms with different number of 
external particles with spin.  Thus, for a theory containing a scalar 
field, $\phi$, and a vector field, $A^\mu$, this expansion would look 
like
$$ \eqalign{
   \Gamma(\Phi_c) &= \int d^4x \left\{ - V(\varphi_c)
    +{1 \over 2} \partial_\mu \Phi_c(x) \partial^\mu \Phi_c(x) Z(\varphi_c)
   \right.\cr & \left.\quad + \partial_\mu \Phi_c(x) A^\mu_c(x) \Phi_c(x) G(\varphi_c)
    + \Phi_c(x)^2  {A^\mu_c}^2 H(\varphi_c)
     + A_{\mu c} A^\mu_c K(\varphi_c) +\cdots \right\}  \, .
  }\eqno(2.8)
$$
Note that the coefficients in this expansion are functions, not functionals. 
The first term, $V(\varphi_c)$, is called the effective potential; it is
equal to the sum of all Feynman diagrams with only external scalar lines and 
with vanishing external momenta.  Those diagrams without loops correspond to
the interactions in the Lagrangian, while those with loops correspond to the 
quantum corrections.

Now we wish to see how spontaneous symmetry breaking is described in this
formalism.  We assume that the Lagrangian possesses a symmetry which is
spontaneously broken if some field acquires a non-zero vacuum expectation 
value.  From Eq.~(2.4) we see that if the external sources vanish, the 
classical field is just the vacuum expectation value of the quantized 
field; this observation, plus Eq.~(2.6), tells us that the condition 
for spontaneous symmetry breaking is
$$
    {\delta \Gamma(\Phi_c(x)) \over \delta\Phi_c(x) } = 0 \, ,
   \quad {\rm for~}\Phi_c(x) \ne 0   \, .
   \eqno (2.9)
$$
Since spontaneous breaking of Poincar\'e invariance does not seem 
to be a physically significant phenomenon, we will restrict ourselves to 
theories in which the vacuum is invariant under translation.  If the 
Lagrangian and the quantization procedure are also Poincar\'e invariant,
then the vacuum expectation value of the fields will be constant in
space-time, and Eq.~(2.9) reduces to 
$$   
    { dV(\varphi_c) \over d \varphi_c} = 0 \, ,
  \quad {\rm for~}\varphi_c \ne 0   \, .
   \eqno(2.10)
$$

Thus we have obtained, as promised, a c-number function whose minimum 
determines the nature of the ground state.  In fact, the effective potential
can give us much more information.  If we compare Eqs.~(2.7) and (2.8), we 
see that the $n$th derivative of $V(\varphi_c)$, evaluated at $\varphi_c=0$, 
is just the $n$-point IPI Green's functions, evaluated at zero external momenta.
Of particular interest is the 2-point IPI Green's function, which is the 
inverse propagator; its value at zero momentum may be taken as the definition 
of the mass.  (It is not exactly the position of the pole in the propagator, 
but we usually expect it to be close.)  Thus, in a theory without spontaneous
symmetry breaking, we may write the scalar meson mass matrix as
$$
     m^2_{ij} = 
    \left.{\partial^2 V \over \partial \varphi_i \partial \varphi_j}
    \right|_{\varphi_c=0}   \, .
   \eqno(2.11)
$$ 
If spontaneous symmetry breaking does occur, then Eq.~(2.11) no
longer holds, since the mass is defined for an isolated particle; that
is, the inverse propagator must be evaluated in the ground state.  To
do this, we make the usual redefinition of the scalar fields; if
$\langle \varphi \rangle$ is the vacuum expectation value of $\phi$,
we define a new quantum field, $\phi'$, and a new classical field,
$\Phi'_c$, by
$$
    \phi'(x) = \phi(x) - \langle \varphi \rangle
   \eqno(2.12)
$$
and 
$$
    \Phi'(x) = \Phi(x) - \langle \varphi \rangle   \, .
   \eqno(2.13)
$$
The mass matrix is then given by
$$
     m^2_{ij} =
    \left.{\partial^2 V \over \partial \varphi'_{ic} 
      \partial \varphi'_{jc}}
    \right|_{\varphi'_c=0}
    = \left.{\partial^2 V \over \partial \varphi_{ic} 
        \partial \varphi_{jc}}
    \right|_{\varphi_c= \langle \varphi \rangle}
    \, .
   \eqno(2.14)
$$ 

There is an additional advantage of the functional formalism, which we
will not make use of, but which is worth pointing out: Instead of
coupling the external sources in Eq.~(2.1) to the elementary fields of
the theory, we could have coupled them to more complicated fields.  We
would then have obtained a set of classical fields which would not
have simple relationships to the elementary quantum fields of the
theory.  This might be desirable in at least two different cases.
First, it allows one to study the possibility of spontaneous symmetry
breaking in a theory without elementary scalar fields, by having a
compound field, rather than an elementary field, acquire a non-zero
vacuum expectation value.  Second, it has the possibility of
describing strong interaction effects in terms of the phenomenological
fields, even though these fields quite likely cannot be simply
expressed in terms of the fundamental quantum fields.

Thus the study of spontaneous symmetry breaking is reduced to the
calculation of the effective potential.  Unfortunately, this is not so
simply done; even if one accepts the validity of a calculation to all
orders of perturbation theory, such a calculation requires summing an
infinite number of Feynman diagrams, which is no mean feat.  It is
therefore necessary to find an approximation for the effective
potential which will require the summation of only some subset of the
relevant Feynman diagrams.

The first expansion to come to mind is that which is most often used
in perturbation theory: expansion in powers of the coupling constants.
However, there are two objections to using this method for the problem
at hand.  First, in theories with spontaneous symmetry breaking one
commonly defines shifted fields, with a corresponding redefinition of
the coupling constants.  This can lead to confusion in comparing the
order of diagrams calculated using different shifts.  But this can be
unraveled if one is careful; it certainly does not invalidate the
method.  The second objection is more basic: We will often be
considering theories which appear to contain massless particles, for
example, massless scalar electrodynamics.   Two diagrams which 
contribute to the effective potential in scalar electrodynamics\
are shown in Fig.~1.  The first is of order $e^4$, while the second is 
of order $e^{24}$.  An expansion in powers of coupling constants assumes
that the second is negligible in comparison with the first, while in fact it is 
much more infrared divergent, and not at all negligible.  This is not merely 
a fluke which occurs because we happen to have considered a theory with 
massless particles.  Spontaneous symmetry breaking, like critical 
phenomena in many-body theory, is associated with correlations between 
widely separated points; the presence of massless particles, which can
mediated long-range forces, is obviously of extreme importance.  In fact, if one 
tried to study  massless scalar electrodynamics and neglected diagrams such
as that in Fig.~1b, one would obtain qualitatively different results.

A better approximation is the expansion by the number of loops in a 
diagram$^5$.  First, let us show that such an expansion is unaffected 
by a shift of the fields.  We introduce a parameter $b$ into the 
Lagrangian by writing
$$
  {\cal L}(\phi_i, \partial^\mu \phi_i; b)
    = b^{-1} 
  {\cal L}(\phi_i, \partial^\mu \phi_i)  \, .
   \eqno(2.15)
$$
The power, $P$, of $b$ associated with a particular diagram will be
$$
   P= I-V \, ,
    \eqno(2.16)
$$ 
where $I$ is the number of internal lines and $V$ is the number of
vertices, since the vertices are obtained from the interaction 
Lagrangian while the propagators are obtained from the inverse of 
the free Lagrangian.  The number of loops, $L$, is equal to the 
number of independent internal momenta.  This is equal to the 
number of momenta ($I$), less the number of energy-momentum delta
functions ($V$), but not counting the delta function corresponding to 
overall energy-momentum conservation.  In other words
$$
   L = I -V +1 = P +1 \, .
  \eqno(2.17)
$$
We see that the number of loops is determined by the power of a 
quantity which multiplies the whole Lagrangian, and does not depend
on the details of how the Lagrangian was written; thus the loop expansion
is unaffected by a redefinition of the fields.

It still remains to be seen whether the loop expansion is a good
approximation.  Certainly the appearance of high powers of $b$ does
not make a diagram small, since $b$ must be set equal to one.
However, many-loop diagrams must contain many powers of the coupling
constants, so the loop expansion is at least as good as the coupling
constant expansion.  Also, diagrams such as that in Fig.~1b are
included in the same order as that in Fig.1a, so the second objection
to the coupling constant expansion is avoided.  (One may ask why
certain two-loop graphs, such as that in Fig.~2, are any less
important than that in Fig.~1b; the answer will appear presently.)  We
shall see as we go on that logarithms will appear in our calculations,
and that the validity of the loop expansion will require not only that
the coupling constants, but also the logarithms, be small.
(Similarly, logarithms appear in the usual applications of the
coupling constant expansion, where one expects the results to hold
only when the logarithms are small.)

Finally, we note that the zero-loop (tree) approximation is equivalent to 
treating the theory classically.

At this point it is probably most instructive to consider a specific
model in order to see how our methods work.

\vfill \eject

\centerline{III.  A Simple Example}

\bigskip

Let us consider a simple, but illustrative, model: the theory of a single
quartically self-coupled scalar field.  The Lagrangian for the theory 
is $^6$
$$
  {\cal L} = {1 \over 2} (\partial_\mu \phi)^2 - {1\over 2}m^2\phi^2
      -{\lambda\over 4!}\phi^4
      +{A \over 2} (\partial_\mu \phi)^2 - {B\over 2}\phi^2
      -{C\over 4!}\phi^4   \, ,
  \eqno(3.1)
$$
where the last three terms are counter-terms to be determined, order
by order in the loop parameter, by the renormalization conditions.
The coupling constant, $\lambda$, may be either positive or negative.
Ordinarily one forbids negative $\lambda$ on the grounds that it leads
to a potential which has no lower bound, but of course this is a
statement only about the zero-loop approximation to the effective
potential; the loop contributions may (or may not) be such as to put a
lower bound on the effective potential, even for negative $\lambda$.

The theory contains a discrete symmetry, namely $\phi \rightarrow -
\phi$.  The conventional wisdom (for positive $\lambda$) is that the
vacuum is symmetric if $m^2$ is positive, and that spontaneous
symmetry breaking occurs if $m^2$ is negative.  (Goldstone's theorem
does not apply, since there is no continuous symmetry.)  However, the
conventional wisdom does not say very much about the case where $m^2$
vanishes, except to warn of the terrible infrared divergences which
may arise.  In fact, this case is not well defined until the
renormalization conditions are specified.  Let us choose the mass
renormalization so that
$$
   \left. {d^2V \over d\phi_c^2} \right|_{\varphi_c =0} =m^2 =0 \, .
   \eqno(3.2)
$$
If the vacuum occurs when $\varphi_c=0$, this means that the inverse
propagator at zero external momentum vanishes; in other words, the
theory really contains a massless particle.  On the other hand, if
spontaneous symmetry breaking occurs, this is not a condition on the
physical inverse propagator and there is not necessarily a massless
particle; one might ask why this should be considered a special case.
The reason is that the point $\varphi_c=0$ is a symmetric one, and the
symmetric formulation of a theory is significant; for example, the 
unitary gauge of
a Higgs theory may be useful for calculating amplitudes, but the
renormalizable gauge is certainly better for understanding the
underlying symmetries of the theory.  In any case, let us consider the
theory with $m^2=0$, and see whether spontaneous symmetry breaking
occurs.  (We still have not completely specified the theory, since the
wave function and coupling constant renormalization conditions have
not been stated, but these are best left for later, when their
motivation will be more apparent.

One last task remains before we can begin calculating the effective
potential: we must learn how to count $i$'s, minus signs, and
factorials.  A term $g\phi^n$ in the Lagrangian leads to a Feynman
diagram with $n$ external $\phi$ lines and a value of $ig(n!)$.
Remembering that the Lagrangian contains the negative of the
potential, we see that the contribution to the effective potential
from a diagram with $n$ external lines is 
$$
   {i \over n!} \varphi^n_c \times ({\rm diagram}) \, .
   \eqno(3.3)
$$

In the zero-loop approximation only one diagram, that shown in Fig.~3,
contributes to the effective potential.  Its contribution is, of
course,
$$ 
   V_{\rm zero-loop} = {\lambda \over 4!} \varphi^4_c \, ,
   \eqno(3.4)
$$
In the next order we have the infinite series of diagrams shown in 
Fig.~4a, as well as the diagrams of Fig.~4b, which arise from the 
one-loop mass and coupling constant counter-terms.   The latter 
clearly do not contain any loops, but they are to be counted here
because we are really expanding in powers of the loop-counting
parameter, $b$, and these terms are of order $b^0$.

In calculating the contribution from the diagrams of Fig.~4a, we must
keep track of certain combinatoric factors.  To understand these, let
the external lines have small non-zero momenta, $\epsilon_1, 
\epsilon_2, \dots, \epsilon_{2n}$.  (Ultimately, we will let these
momenta go to zero.)  We now count the one-loop diagrams with 
$2n$ external lines.  First, we consider the case where $n \ge 3$.
We note the following:

1) There are $(2n)!$ ways of arranging the external momenta.

2) Interchanging the external moment at a vertex does not give a
new diagram, so we have a factor of $({1\over 2})^n$.

3) Rotating or reflecting a diagram does not give a new diagram, so 
we have another factor of $1/(2n)$.

The contributions from one-loop diagrams with $2n$ external lines
($n \ge 3$) is thus
$$
 \left[(2n!) \left({1\over 2}\right)^n {1 \over 2n} \right]
  \left[{i \over (2n)!} \varphi^{2n}_c\right]({\rm diagram})
   = {i \over 2n} \left({\varphi_c^2\over 2}\right)^n  
   ({\rm diagram}) \, .
\eqno(3.5)
$$

If $n$ equals 1 or 2, point (3) must be modified.  For $n=1$, the diagram
cannot be reflected or rotated; for $n=2$, reflection and rotation are
the same, 
so there is only a factor of $1/2$, rather
than $1/4$.  However, in both of these cases the Feynman rules include an 
extra factor of $1/2$.  

We can now write an expression for the one-loop approximation to the 
effective potential:
$$
   V_{\rm one-loop~approx} = {\lambda \over 4!} \varphi_c^4
     + \sum_{n=1}^\infty \int {d^4 k \over (2\pi)^4}
     \left[{i \over 2n} 
    \left( {\lambda\varphi^2_c/2 \over k^2 +i\epsilon}\right)^n\right]
     + {B^{(1)}\over 2} \varphi^2_c + {C^{(1)}\over 4!} \varphi^4_c
  \, .
\eqno(3.6)
$$
We notice that the integrals for $n \ge 2$ are infrared divergent.  We also 
notice that the infinite sum can be done, yielding
$$
   V_{\rm one-loop~approx} = {\lambda \over 4!} \varphi_c^4
     + {i \over 2}\int {d^4 k \over (2\pi)^4}
     \log\left(1 - {\lambda\varphi^2_c/2 \over k^2 +i\epsilon}\right)
     + {B^{(1)}\over 2} \varphi^2_c + {C^{(1)}\over 4!} \varphi^4_c
  \, .
\eqno(3.7)
$$
The infrared divergence has disappeared, being replaced by a logarithmic
singularity at the origin of classical field space.  It is not too
surprising that this should occur, since we expect infrared divergences 
only if the vacuum is at $\varphi_c =0$. If spontaneous symmetry
breaking occurs, then $\phi$ does not remain massless, and there is no
reason to expect infrared problems.

We do the integral by rotating into Euclidean space and cutting off the 
integral at $k^2 = \Lambda^2$.  We obtain
$$ \eqalign{
   V_{\rm one-loop~approx} &= {\lambda \over 4!} \varphi_c^4
     + {1 \over 64\pi^2} \left\{ \lambda \varphi_c^2 \Lambda^2
     + {\lambda\varphi_c^4 \over 4}
        \left[\log {\lambda \varphi_c^2\over 2 \Lambda^2} -{1 \over 2}
        \right] \right\}   \cr &\quad
     + {B^{(1)}\over 2} \varphi^2_c + {C^{(1)}\over 4!} \varphi^4_c
  \, .  
}\eqno(3.8)
$$

We must now determine the value of the renormalization counter-terms.  
The mass renormalization condition was given in Eq.~(3.2).  From it
we deduce that 
$$    
  B^{(1)} = - {\lambda \Lambda^2 \over 32\pi^2} \, .
\eqno(3.9)
$$
Now we come to the coupling constant renormalization condition, which we
have left unspecified so far.  Two possibilities are 
$$
    \left. {d^4V \over d\varphi_c^4} \right|_{\varphi_c=0} = \lambda
\eqno(3.10a)
$$
and 
$$
    \left. {4! \over \varphi_c^4}V \right|_{\varphi_c=0} = \lambda \, .
\eqno(3.10b)
$$
Unfortunately, neither of these is acceptable, because of the logarithmic 
singularity at the origin of the classical field space.  Instead, we 
choose an arbitrary value of $\varphi_c$, which we denote by $M$, and require
either 
$$
    \left. {d^4V \over d\varphi_c^4} \right|_{\varphi_c=M} = \lambda
\eqno(3.11a)
$$
or
$$
    \left. {4! \over \varphi_c^4}V \right|_{\varphi_c=M} = \lambda
     \, .
\eqno(3.11b)
$$
It is important to remember that the choice of $M$ is arbitrary; if a 
different value is chosen for $M$, one obtains a different value for
$\lambda$, but the theory remains unchanged.  In other words, as $M$ is 
varied $\lambda$ traces out a curve $\lambda(M)$.  Although two 
parameters appear, there is actually only a one-parameter family of 
theories, one theory for each curve.  We shall return to this point 
later, when we discuss the use of the renormalization group.  (Note that
this is analogous to what happens in the usual treatment of theories with
massless particles.  Because of the infrared divergences, some of the 
renormalizations cannot be done at zero external momentum; instead,
they are done at an arbitrary non-zero momentum.)

For the moment, let us choose Eq.~(3.11a) as the renormalization 
condition, since it is closer to the definition of the physical 
4-point IPI Green's function. (It is, in fact, the definition of the
physical Green's function if $M$ is equal to the vacuum expectation 
value of $\phi$.)  We find
$$
   C^{(1)} = -{3 \lambda^2 \over 32 \pi^2} 
       \left(\log{\lambda M^2 \over 2\Lambda^2} +{11\over 3} \right)
\eqno(3.12) 
$$
and 
$$
   V_{\rm one-loop~approx} = {\lambda \over 4!} \varphi_c^4
   + {\lambda^2 \varphi_c^4 \over 256\pi^2} 
      \left[\log {\varphi_c^2 \over M^2} - {25\over 6}\right] \, .
\eqno(3.13)
$$
We note several things about our final expression for the effective
potential:

1) $V$ still contains the logarithmic singularity in $\phi_c$, but the
singularity at $\lambda=0$ has disappeared.

2) $M$ is in fact arbitrary; if we let $M$ go to $M'$, we can choose
a new coupling constant, 
$$
    \lambda' =   \left. {d^4V \over d\varphi_c^4} \right|_{\varphi_c=M'}
    = \lambda + {3 \lambda^2 \over 32 \pi^2} \log {{M'}^2\over M^2} \, ,
\eqno(3.14)
$$ 
and find 
$$
   V = {\lambda' \over 4!} \varphi_c^4
   + {{\lambda'}^2 \varphi_c^4 \over 256\pi^2}
      \left[\log {\varphi_c^2 \over {M'}^2} - {25\over 6}\right] \, .
\eqno(3.15)
$$
This, to the order of approximation to which we are working, describes
the same theory as Eq.~(3.13)

3) The behavior of $V$ is shown in Fig.~5.  Since the logarithm of 
a small number is negative, it appears that there is a maximum at the 
origin and a minimum at a non-zero value of $\varphi_c$.  
Differentiating $V$, we find 
$$ 
   {dV \over d\varphi_c} = {\lambda\over 6} \varphi_c^3
      + {\lambda^2 \varphi_c^3 \over 64\pi^2} 
    \left[\log {\varphi_c^2 \over M^2} - {11\over 3}\right] \, .
\eqno(3.16)
$$
Thus the minimum occurs at a value of $\varphi_c$ determined by 
$$  
  \lambda \log{\pc^2\over M^2} -{11\lambda \over 3} = -{32\pi^2 \over 3}
    \, .
\eqno(3.17)
$$
Thus it would appear that this theory exhibits spontaneous symmetry
breaking.  However, we must consider the effects of many-loop
diagrams, and see whether there are solutions of Eq.~(3.17) which are
consistent with the validity of the one-loop approximation.

It turns out that increasing the number of loops brings in not only 
additional powers of $\lambda$, but also additional logarithms; that 
is, the leading behavior of the $n$-loop contribution to the 
effective potential is of the order of 
$$  
    \lambda \left(\lambda \log{\pc^2\over M^2} \right)^n \, .
\eqno(3.18)
$$
Thus the one-loop approximation can be valid only if both $|\lambda|$ and
$|\lambda \log(\pc^2/M^2)|$ are small.  However, it is clear that 
Eq.~(3.17) cannot be satisfied unless one of these is large.  There may 
be a minimum of the effective potential away from the origin, but if there 
is, it is in a region where our calculational methods fail.  Similarly, 
we cannot trust the prediction that there is a maximum at the origin.  
Later we shall see that by the use of the methods of the renormalization 
group we can improve the situation, but for the present we simply do
not know whether spontaneous symmetry
breaking occurs.

Of course, we should not be too surprised that our approximation did
not give us any reliable evidence of spontaneous symmetry breaking.
The zero-loop approximation to the effective potential indicates that
spontaneous symmetry breaking does not occur.  If we want the one-loop
approximation to change this, we should expect that it must be of the
same order of magnitude as the zero-loop term; this is possible only
if the interaction is strong, in which case a perturbative methods is
expected to fail.  At this point, we might conclude that our methods
will never indicate spontaneous symmetry breaking unless it is already
predicted by the classical approximation; however, this is not quite
true.

Consider a theory with two interactions, A and B.  Suppose that only A
contributes to the effective potential in the zero-loop approximation,
but that B contributes through loop diagrams.  Even if A and B are
both weak, it may be possible to make the one-loop B contribution and
the zero-loop A contribution be of the same order of magnitude by
adjusting the relative strength of A and B.  If so, the one-loop
contributions might give rise to spontaneous symmetry breaking, not
because they are large, but because they involve a new type of
interaction.  Since both interactions are weak, and there are no other
interactions to be introduced, the contributions from diagrams with
two or more loops should be small; therefore, the one-loop
approximation should be valid.

One might ask whether it is possible to construct a theory in which
the loop expansion is valid, but in which the diagrams with two or
more loops introduce new interactions which qualitatively change the
effective potential; we not found any.  To see the difficulty,
consider the electrodynamics of a scalar with charge $e$ and a fermion
with charge $g$, with an additional quartic scalar self-coupling with
coupling constant $\lambda$.  The zero-loop approximation to the
effective potential will be of the order of $\lambda$, while the
one-loop contributions will be of the order of $\lambda^2$ or $e^4$.
The fermion-photon interaction first contributes in the two-loop
approximation, through corrections to the photon propagator.  These
contributions will be of the order of $g^2e^4$; for these to be
significant in comparison with the one-loop terms, $g^2$ must be
large, in which case the perturbation theory is no longer valid.

Let us now justify Eq.~(3.18) by deriving some simple rules which make it 
possible to sum all the graphs of any particular order as easily as 
we summed all of the one-loop graphs.  Consider, for example, the two-loop
graphs of the type shown in Fig.~6a.  To obtain the contribution of 
these graphs to the effective potential, we must sum over all values 
of $l$, $m$, and $n$ from zero to infinity.  In doing the sum, we 
must remember the following factors:

1) A factor of $(2l + 2m +2n)!$, which cancels the factorial in 
Eq.~(3.3), just as in the one-loop calculation.

2)  A factor of $(1/2)^{l+m+n}$, from interchanging the lines at each of 
the vertices with two external lines; this also is analogous to 
the one-loop calculation.

3)  A factor of $1/(3!)$, arising from the symmetry of the diagram.

We can do the sums over $l$, $m$, and $n$ independently.  Each sum is
of the form
$$
   \sum_{j=0}^\infty \left(-{i\lambda\pc^2 \over 2}\right)^j
       \left( {i \over k^2 +i\epsilon} \right)^{j+1} 
     = {i \over k^2 - {1\over 2}\lambda\pc^2 +i\epsilon}
  \, .  \eqno(3.19)
$$
In addition we have a factor of $-i \lambda \pc$ at each of the two
remaining vertices and the factor of $1/6$ from point (3).  This is
just what we would have obtained from the graph of fig.~6b if the
$\phi$ had a mass of ${1 \over 2}\lambda\pc^2$.  (The Feynman rules
would include the $1/6$, again from the symmetry of the graph.)  We
shall call those graphs which, like that in Fig.~6b, have no vertices
with  two external lines prototype graphs.  It is clear that the
above discussion holds for all prototype graphs.  We thus obtain a
rule for calculating the $n$-loop contribution to the effective
potential: Draw all of the $n$-loop prototype graphs and calculate
them using ordinary Feynman rules, except with a propagator given by
Eq.~(3.19).  The only exception to this rule is the case of one-loop
graphs, which alone are invariant under rotation.  We note that the
prototype graph in this case is a circle, which does not correspond to
any Feynman diagram arising from ordinary perturbation theory.

All that remains is to show that there are a finite number of $n$-loop
prototype graphs.  We define a type-$n$ vertex to be one with $n$
internal lines, and denote the number of type-$n$ vertices in a graph
by $V_n$.  By definition, an IPI graph does not contain any type-1
vertices, while a prototype graph contains only type-3 and type-4
vertices.  For an IPI graph,
$$
   V = V_2 +V_3 +V_4 \, .
\eqno(3.20)
$$
Since every internal line has two ends, we see that 
$$
   2I = 2V_2 +3V_3 +4V_4 \, .
\eqno(3.21)
$$
Thus, 
$$  
     L = I-V+1 = {1\over 2}V_3 +V_4 +1   \, .
\eqno(3.22)
$$ 
Thus, for a given $L$, only a finite number of type-3 and type-4
vertices are allowed.  Since only a finite number of graphs can 
be made with a given finite number of vertices, the number of 
prototype graphs in any order is finite.

A few words of caution:

1) One must not forget diagrams arising from the insertion of
counter-terms.  These are calculated by drawing prototype graphs with
counterterm insertions drawn explicitly, remembering that the order of
the counter-term must be included in determining the order of a
diagram.  The number of such prototype graphs in any order is finite,
since all counter-terms are at least one-loop in order.

2)  The factor of ${1\over 2} \lambda \pc^2$ in Eq.~(3.19) looks like
a mass, but it is not; it is not even a constant.  However, it is
related to the mass if $\pc$ is set equal to the vacuum expectation 
value of $\phi$.

3) One must remember that we are summing all IPI graphs, and that the
use of prototype graphs is merely a device to perform the sum.  Thus
we must include the prototype graph in Fig.~7, even though it has no
external lines.  The reason is that this prototype graph stands for a
sum of graphs which do have external lines.  However, it also includes
the graph which looks exactly like it and which does not have any
external lines; this graph must be subtracted after the prototype
graph is calculated.  That is, the total contribution of this
prototype graph to the effective potential is
$$  
   i (-i \lambda) \left\{\left[ {1 \over 2} \int{d^4k\over (2\pi)^4}
     {i \over k^2 -{1\over 2}\lambda \pc^2 +i\epsilon}\right]^2
    - \left[ {1 \over 2} \int{d^4k\over (2\pi)^4}
     {i \over k^2  +i\epsilon}\right]^2 \right\} \, .
\eqno(3.23)
$$

We can now calculate some many-loop graphs, and verify Eq.~(3.18). It
is now also possible to specify the wave function renormalization condition,
which we have not yet done.  This is a condition on the term in the 
effective action which is of the form 
$$
    \partial_\mu \Phi_c \partial^\mu \Phi_c  Z(\pc) \, .
\eqno(3.24)
$$

In the zero-loop approximation, $Z$ will be 1.  The one-loop contribution to 
$Z$ arises from the counter-term and from the one-loop diagrams which 
have two external $\phi$'s with four-momenta $p$ and $-p$, and any 
number of external $\phi$'s with vanishing four-momenta.  (Actually, it
is just the term proportional to $p^2$ in the Taylor series expansion of
these diagrams.)  The sum of these diagrams can be calculated using the 
methods outlined above:  one draws prototype graphs in which the vertices
with non-vanishing external momenta are shown explicitly, while the vertices
with two zero-momentum external lines are implicit in the propagators.  The
resulting integrals have ultra-violet divergences and must be cut off; 
a counter-term is necessary to make $Z$ cutoff-independent.  The most
natural choice for the renormalization condition is 
$$
    Z(0) = 1  \, .
\eqno(3.25)
$$

Unfortunately, the same difficulty arises here as arose with the
coupling constant renormalization: $Z$ contains a logarithmic
singularity at $\pc=0$.  
Thus, we must renormalize at some arbitrary non-zero value of $\pc$.
For simplicity, we choose the value we used
for the coupling constant renormalization, and have 
$$
    Z(M) = 1  \, .
\eqno(3.26)
$$

Again we see the similarity to the usual method of renormalization in
momentum-space.  There, the wave function renormalization, like the
coupling constant renormalization, would have an infrared divergence,
while the mass renormalization would not; in our method, the first two
have logarithmic singularities, while the third does not.

\vfill \eject

\centerline{IV. Bigger and Better Models}

\bigskip

In this chapter we shall extend the methods of the previous chapter to 
models which include more than one field; we shall continue, however, to
restrict ourselves to models in which no mass term appears in the 
Lagrangian.  (As mentioned previously, this does not necessarily imply
that the theory contains massless particles.)   We shall also restrict
ourselves to renormalizable theories, since only for these do we have
any assurance that higher order calculations will be finite.

In a theory with many quantized fields, we find it useful to define a
classical field corresponding to each quantized field; the effective
action is a functional of all these classical fields.  However, in
expanding $\Gamma$ in Eq.~(2.8), we defined the effective potential to
depend only on the spinless classical fields.  The reason is that we
are only interested in cases where the vacuum is Lorentz invariant;
therefore we want the vacuum expectation value of any field with spin
to be zero.

Just as with the theory of a single scalar field, it is possible to derive
rules which enable us to sum the infinite number of Feynman diagrams 
which occur in each order in the loop expansion.  We begin by considering
graphs that contain only spin-zero particles.  We arrange the real
spinless fields in a vector $\vec\phi$, with components $\phi_i$.
(One should not be misled by the notation to conclude that these fields
necessarily belong to a representation of some symmetry group.)  Also,
we define $V_0(\vec\pc)$ to be the zero-loop approximation to the 
effective potential.  (It is, of course, of the same form as the potential
terms in the Lagrangian, except that it is a function of the classical,
rather than of the quantized, fields.)  Finally, we define a matrix
$U(\vec\pc)$ by 
$$
    U_{ij}(\vec\pc) = {\partial V_0(\vec\pc) \over 
          \partial \varphi_{ic} \partial \varphi_{jc} } \, .
\eqno(4.1)
$$
This is exactly the factor which occurs at each scalar vertex with two
external lines.  We can now calculate the sum of all one-loop diagrams
with only spinless particles; by comparison with Eq.~(3.8), we see that
the result is 
$$
    V_{\rm spin-0~loop} = {1 \over 64\pi^2} 
           {\rm Tr}\,\left\{2U(\vec\varphi) \Lambda^2
              + U(\vec\varphi)^2 
   \left[ \log {U(\vec\varphi) \over \Lambda^2} - {1 \over 2} 
        \right] \right\} \, .
\eqno(4.2)
$$
We note that $U$ is a symmetric matrix and can therefore the diagonalized;
there is thus no difficulty in interpreting this expression.  In order
to calculate sums of multi-loop graphs, we determine Feynman rules for 
prototype graphs.  The propagator will now be a matrix; referring to 
Eq.~(3.19), we see that it is given by 
$$
   \Delta_{ij}(\vec\pc) = \left[ {i\over k^2 -U(\vec\pc) 
      +i \epsilon} \right]_{ij} \, .
\eqno(4.3)
$$

Next we consider the effect of the spin-$1 \over 2$ fields, which 
we arrange in a vector $\vec \psi$.  The Yukawa coupling term in the
Lagrangian is of the form
$$
  {\cal L}_{\rm Yuk} = \bar\psi_i F_{ij}(\vec\pc) \psi_j
      = \bar\psi_i \left[A_{ij}(\vec\pc) I + B_{ij}(\vec\pc)i\gamma^5
      \right] \psi_j \, ,
\eqno(4.4)
$$
where both $A$ and $B$ are Hermitian matrices.  In computing the 
contribution to the effective potential from diagrams with a 
single fermion loop, the matrix $F(\vec\pc)$ will appear at each 
vertex.  In summing these diagrams one must remember that one-loop
diagrams with an odd number of vertices vanish, since the trace of
an odd number of Dirac matrices is zero.  Thus, it is appropriate
to group the vertices in pairs, noting that
$$ 
    {1 \over \not{\! k}} F(\vec\pc)
    {1 \over \not{\! k}} F(\vec\pc) = {1 \over k^2} 
       F(\vec\pc) F(\vec\pc)^* \, .
\eqno(4.5)
$$
The sum also differ from that for the scalar loops in that there
is the usual minus sign for a fermion loop and in that,
since fermion lines are directed, there is no factor of $1/2$
arising from the reflection symmetry of the diagram.  The latter
factor is compensated for by another factor of $1/2$ from 
summing only the even terms in the sum.  Thus, the contribution 
of the fermion one-loop diagrams is 
$$ \eqalign{
   V_{\rm spin-1/2~loop} &= - {1 \over 64\pi^2}
           {\rm Tr}\,\left\{2 \Lambda^2 F(\vec\pc) F(\vec\pc)^*
     \right. \cr &\quad \left.
        + \left[F(\vec\pc) F(\vec\pc)^*\right]^2
   \left[ \log { F(\vec\pc) F(\vec\pc)^* \over \Lambda^2} - {1 \over 2}
        \right] \right\} \, ,
}\eqno(4.6)
$$
where the trace is over both Dirac and particle indices.  

In determining the fermion propagator to be used in prototype 
graphs, both even and odd numbers of vertices must be included in the
sum, since no trace is being taken.  One obtains
$$
    S_{ij}(\vec\pc) = \left[ { i\over \not{\! k}- F(\vec\pc)
           }\right]_{ij} \, .
\eqno(4.7)
$$

Lastly we come to the vector particles, which we will arrange in a 
vector $\vec{A^\mu}$; in the cases we will consider, these will 
always be associated with gauge symmetries, so that the theory will
be renormalizable.  Here matters are complicated by the fact that 
in gauge theories there are two types of scalar-vector vertices, 
shown in Fig.~8.  Because of the vertex of Fig.~8a, there are 
diagrams, such as that in Fig.~9, which have both a scalar and a
vector particle going around the loop.  However, these diagrams will
not contribute to the effective potential if we work in the 
Landau gauge, where the vector propagator is transverse.  
Because the external scalars have zero momentum, the internal
scalars and the vectors have the same momentum; thus, when the 
Landau gauge propagator is multiplied by the momentum 
factor from the vertex the result is zero.   We will do all of 
our calculations in the Landau gauge.

We define a matric $G(\vec\pc)$ to describe the couplings giving
rise to the vertex of Fig.~8b.  It appear in the Lagrangian as
$$ 
  {\cal L} = \cdots + {1\over 2} A^\mu_i A_{\mu j} G_{ij}(\vec\pc)
     + \cdots  \, .
\eqno(4.8)
$$   
$G$ is also given in terms of the infinitesimal generators, $T_i$,
of the gauge group:
$$
    G_{ij}(\vec\pc) = g_i g_j (T_i \vec\varphi, T_j \vec\varphi) \, .
\eqno(4.9)
$$
Here $g_i$ and $g_j$ are the appropriate gauge coupling constants.

The computation of the sum of one-loop diagrams with a gauge 
particle going around the loop can now be done.  The only new 
complication is the presence of the numerator factors in the 
vector propagator.  In the Landau gauge these work out quite 
easily; their only effect is to multiply the result by a factor 
of 3.  Thus, we have 
$$  V_{\rm gauge~loop} =  {3 \over 64\pi^2}
           {\rm Tr}\,\left\{2 \Lambda^2 G(\vec\pc)
              + G(\vec\pc)^2
   \left[ \log {G(\vec\pc) \over \Lambda^2} - {1 \over 2}
        \right] \right\} \, .
\eqno(4.10)
$$

The calculation of the vector meson propagator to be used in 
prototype graphs is also straightforward; one obtains
$$
   D^{\mu\nu}_{ij} = \left[ {i\left(-g^{\mu\nu} + {k^\mu k^\nu \over k^2}
     \right)
       \over k^2 - G(\vec\pc) + i\epsilon} \right]_{ij} \, .
\eqno(4.11)
$$

If the vector particles are associated with a non-Abelian gauge group,
the theory will also involve fictitious ghost particles$^7$.  If one
works in the Landau gauge, the ghosts do not couple to the other
scalar particles and therefore need no further consideration at this
point.  There will be higher-order graphs contributing to the effective
potential which have ghosts coupled to the gauge particles, but in these
the ghost part of the diagram is calculated as usual.  We will see, however,
that the ghosts will complicate the renormalization procedure.

We will not discuss the renormalization procedure in detail at this point,
since it is similar to that for the theory of Chapter~III; when appropriate, 
we will make some comments in the context of particular theories.  Let us
now proceed to consider several models, to see if our methods can 
detect spontaneous symmetry breaking in any of them.

\medskip
\centerline{ 1. A Scalar SO($n$) Model}

\medskip

This model is very much like our original simple model; instead of 
a single scalar fields there are $n$ real fields, transforming as a
vector under the group SO($n$).  The Lagrangian is 
$$
  {\cal L} = {1 \over 2} \partial_\mu \vec\phi \cdot 
     \partial^\mu \vec\phi  -  {\lambda\over 4!}
       (\vec\phi \cdot \vec\phi)^2 + {\rm counter{\mathhyphen}terms} \, .
\eqno(4.12)
$$
The effective potential is expected to be a function of the $n$ 
classical fields.  However, we notice that because of the symmetry of 
the theory, the effective potential can only be a function of
$$
    \pc^2 = \sum_{k=1}^n \varphi_{kc}^2 \, .
\eqno(4.13)
$$
Thus we can calculate the effective potential for the case where only
$\varphi_{1c}$ is non-zero, and then immediately extend the results to
the general case.  In other words, it is sufficient to calculate only 
diagrams with external $\phi_1$'s.  (This is just another way of saying 
that the vacuum expectation value of $\vec\phi$ must point in a particular
direction.)  We expect two different types of one-loop diagrams: those
with a $\phi_1$ running around the loop, and those with one of the 
other $\phi_k$'s running around the loop.  Referring to Eqs.~(4.1) and
(4.2), we see that 
$$
     \left. U(\vec\pc) \right|_{\varphi_{1c}\ne 0, 
       \varphi_{2c}=\varphi_{3c}=\cdots \varphi_{nc}=0}   =
   \left(\matrix{ {1\over 2}\lambda \varphi_{1c}^2 & 0 & \cdots & 0 \cr
    0 & {1\over 6}\lambda \varphi_{1c}^2 & & \cr
    \vdots & & \ddots \cr
     0 & & & {1\over 6}\lambda \varphi_{1c}^2 } \right)
\eqno(4.14)
$$
and thus
$$ \eqalign{
   &\left.  V(\vec\pc)\right|_{\varphi_{1c}\ne 0,
       \varphi_{2c}=\varphi_{3c}=\cdots \varphi_{nc}=0} 
    = {\lambda \over 4!} \varphi_{1c}^4 
    +{1 \over 64\pi^2} \left\{ 2\Lambda^2 \left({\lambda\over 2} 
        + {(n-1)\lambda\over 6} \right) \varphi_{1c}^2
   \right.\cr &\qquad\qquad\qquad \left.
    +{\lambda^2 \over 4} \varphi_{1c}^4 \
      \left[\log {\varphi_{1c}^2 \over 2\Lambda^2} -{1 \over 2}
          \right] 
   +{(n-1) \lambda^2 \over 36}
         \left[\log {\varphi_{1c}^2 \over 6 \Lambda^2} -{1 \over 2}
          \right]
      \right\}
   \cr &\qquad\qquad\qquad  +{\rm counter{\mathhyphen}terms} \, .
}\eqno(4.15)
$$
We see that there are, as expected, two types of one-loop contributions.

Our renormalization conditions are similar to those of the theory with a 
single scalar particle, namely
$$
    \left.{\partial^2 V\over \partial \varphi_{1c}^2}
   \right|_{\varphi_{1c}= \cdots \varphi_{nc} =0} =0  \, ,
\eqno(4.16a)
$$
$$
    \left.{\partial^4 V\over \partial \varphi_{1c}^4}
   \right|_{\varphi_{1c}=M, \varphi_{2c}= \cdots \varphi_{nc} =0} =
    \lambda \, ,
\eqno(4.16b)
$$
and 
$$
   Z(\varphi_{1c}=M, \varphi_{2c}= \cdots \varphi_{nc}=0) = 1 \, .
\eqno(4.16c)
$$
Performing the renormalization, we obtain the final expression for the
one-loop approximation to the effective potential:
$$
    V(\vec\varphi) = {\lambda\over 4!} \pc^4 
      +{1 \over 64\pi^2}\left({\lambda^2 \over 4} + 
          {(n-1) \lambda^2 \over 36}\right)\pc^4 
       \left[\log{\pc^2 \over M^2} -{25\over 6} \right] \, .
\eqno(4.17)
$$

This effective potential is of the same form as that of our previous
model, but it differs from the previous one in having the factor of
$n-1$ in the logarithmic term.  Thus, we might hope that for
sufficiently large $n$ this potential will exhibit a minimum within
the range of validity of the one-loop approximation.  (In terms of the
comments of the previous chapter, we can attribute this to the
presence in this theory of two types of interactions: the $\phi_i^4$
interaction and the $\phi_i^2 \phi_j$2 interaction ($i \ne j$).  The
latter first contributes to the effective potential in the one-loop
approximation.)  If we differentiate, we obtain, instead of
Eq.~(3.17),
$$
     \lambda \left(1 + {n-1 \over 9} \right) \left( 
   \log{\pc^2 \over M^2} -{11\over 3} \right) = - {32 \pi^2 \over 3}
    \, .
\eqno(4.18)
$$
Certainly this equation can be satisfied with both $|\lambda|$ and 
$|\log(\pc^2/M^2)|$ being kept small if $n$ is made sufficiently large;
we can obtain a minimum by having many types of loops, rather than by making
a single loop contribution large.  Unfortunately, letting $n$ be large also
invalidates the one-loop approximation.  For example, consider the two-loop
prototype graphs of the form shown in Fig.~7.  Each of the $n$ $\phi_k$'s
can be allowed to run around the left loop and and each of the 
$n$ $\phi_k$'s can be allowed to run around the right loop; thus there are
$n^2$ prototype graphs of this type.  Each higher order will bring in 
another factor of $n$, so the $L$-loop contribution will be of the order of 
$$
    \lambda \left( n\lambda \log {\pc^2 \over M^2} \right)^L \, .
\eqno(4.19)
$$
Thus, if $n$ is large enough to satisfy Eq.~(4.18), it is large enough to
invalidate the one-loop approximation.  (We see that we were mistaken in 
identifying two types of interactions; a better way of describing the 
theory is to say that it contains a single many-component field and only 
one interaction.)  This model is no improvement over our first one.

\medskip
\centerline{2. A Yukawa Model}

\medskip

Next we consider a theory with a single scalar field and a single
fermion field, with the Lagrangian
$$
  {\cal L} = {1 \over 2} \partial_\mu \phi \partial^\mu \phi
     + i \bar \psi \not{\! \partial} \psi 
     + ig \bar \psi \gamma^5 \phi \psi
   + {\lambda \over 4!} \phi^4 + {\rm counter{\mathhyphen}terms} \, .
\eqno(4.20)
$$
(The quartic scalar self-coupling is required for renormalizability.)
Because this theory contains two genuinely distinct interactions, it 
may be possible to adjust matters so that the one-loop corrections 
qualitatively change the nature of the vacuum.  

Because there is only one scalar field and one fermion field, the 
matrices $U$ and $F$ are trivial:
$$
    U = {1 \over 2} \lambda \pc^2 
\eqno(4.21a)
$$
and 
$$
    F = ig \pc \gamma^5  \, .
\eqno(4.21b)
$$
The calculation of the one-loop approximation to the effective 
potential is straightforward; after renormalization one obtains
$$
   V = {\lambda\over 4!} \pc^4 
     +  {1 \over 64\pi^2}\left({\lambda^2 \over 4} -4g^2 \right) \pc^4
          \left[\log{\pc^2 \over M^2} -{25\over 6} \right] \, .
\eqno(4.22)
$$

Indeed, the Yukawa coupling, which first contributes to the effective 
potential in the one-loop approximation, does have an important 
qualitative effect:  The effective potential has the shape of the previous
ones (Fig.~5) only if $g^4$ is less than $\lambda^2/16$; if $g^4$ is
larger than $\lambda^2/16$, the effective potential has the shape 
shown in Fig.~10.  The latter case is clearly unacceptable; the effective
potential has no lower bound as the classical field approaches infinity,
so the vacuum does not exist (unless higher-order contributions make the
effective potential turn upward, but in that case the one-loop 
approximation is still useless).  And the former case has the same 
difficulties as our earlier models, but to a greater degree: the minimum
is even further from the region of validity of the one-loop approximation.
Thus, we have found a theory where the one-loop contribution to the effective
potential has an important qualitative effect; unfortunately, it isn't
the sort of effect we were looking for.

\medskip
\centerline{3.  Massless Scalar Electrodynamics}

\medskip

Now we consider the theory of two real (or one complex) scalar fields
coupled to a gauge vector meson, with the Lagrangian
$$ \eqalign{
  {\cal L} &= {1 \over 4} F^{\mu\nu} F_{\mu\nu} 
     + {1\over 2} \left( \partial_\mu \phi_1 - e A_\mu \phi_2\right)^2
     + {1\over 2} \left( \partial_\mu \phi_2 + e A_\mu \phi_1\right)^2
   - {\lambda\over 4!} (\phi_1^2 +\phi_2^2)^2
    \cr & \quad + {\rm counter{\mathhyphen}terms}  \, ,
}\eqno(4.23)
$$
where
$$  
      F^{\mu\nu} = \partial^\nu A^\mu - \partial^\mu A^\nu  \, .
\eqno(4.24)
$$
If this theory has a symmetric vacuum, it describes the electrodynamics
of a massless charged scalar meson.  If spontaneous symmetry breaking
occurs, then by the usual Higgs mechanism we obtain the theory of a 
single massive scalar meson coupled to a massive vector meson.  To 
determine which is the case is to determine whether the infrared 
divergences associated with massless charged scalar particles are 
strong enough to prevent such particles from existing.

The calculation of the one-loop approximation to the effective
potential is straightforward, as is the renormalization; one obtains
$$
  V = {\lambda\over 4!} \pc^4
    +  {1 \over 64\pi^2}\left({\lambda^2 \over 4} + {\lambda^2 \over 36}
    + 3e^4\right) \pc^4
          \left[\log{\pc^2 \over M^2} -{25\over 6} \right] \, .
\eqno(4.25)
$$
where 
$$
    \pc^2 = \varphi_{1c}^2 + \varphi_{2c}^2 \, .
\eqno(4.26)
$$
This effective potential also has the shape shown in Fig.~5; it
appears to have a minimum at a non-zero value of $\pc$.  We again 
must determine whether the minimum occurs within the range of validity of
the one-loop approximation.  We can simplify our equations if we recall that
$M$ is an arbitrary parameter; we are certainly allowed to choose $M$ to be the
location of the minimum of the effective potential, $\langle \phi \rangle$.
We find
$$
  \left.{dV \over d\pc} \right|_{\pc=M= \vev} = {\lambda\over 6}\vev^3
    + {1 \over 64\pi^2}\left({\lambda^2 \over 4} + {\lambda^2 \over 36}
    + 3e^4\right) \left( -{44\over 3} \vev^2 \right) \, .
\eqno(4.27)
$$
Thus, we must see if we can satisfy
$$
    \lambda = {11\over 8\pi^2} \left( 3e^4 
       + {\lambda^2 \over 4} + {\lambda^2 \over 36}\right)
\eqno(4.28)
$$
while keeping both $e$ and $\lambda$ small enough that our approximation 
is valid.  Clearly we can do this if we choose $\lambda$ to be of the 
order of $e^4$.  If we do this, we should ignore the $\lambda^2$ terms,
since they are of the same order of magnitude as the $e^8$ terms 
we expect from two-loop diagrams.  Thus, in the one-loop
approximation we find that if 
$$
    \lambda ={33 e^4\over 8\pi^2}  \, ,
\eqno(4.29)
$$
there will be a minimum in the effective potential which is within the 
region where the approximation is valid.  The gauge coupling, which 
only contributes to the effective potential through loop diagrams, has 
caused spontaneous symmetry breaking.

If we substitute Eq.~(4.29) back into Eq.~(4.25), we obtain
$$ 
   V = {3e^4 \over 64\pi^2} \pc^4 \left[\log{\pc^2 \over \vev^2} 
     -{1\over 2} \right] \, .
\eqno(4.30)
$$
All reference to the parameter $\lambda$ has disappeared from the effective
potential.  Note that even if we had chosen to define $\lambda$ 
differently (for example, by using Eq.~(3.11b)), we would have obtained
different results for Eqs.~(4.25-29), but the same result for Eq.~(4.30).

At first glance, a very strange thing seems to have occurred: Starting
with two dimensionless parameters, $e$ and $\lambda$, we have obtained
a theory with one dimensionless parameter, $e$, and one dimensional
parameter, $\vev$.  Furthermore, the dependence on $\vev$ is trivial,
being determined solely by dimensional considerations.  A little
thought will make this seem less strange.  Originally, the theory was
really characterized by three parameters: $e$, $\lambda$, and $M$;
however, one two of these were independent and the dependence on $M$
was not explicitly shown.  The final description of the theory is also
characterized by three parameters: $e$, $\vev$, and $\vev/M$; again
only two are truly independent, and we have not shown the dependence
on the third parameter,\ which we have set equal to one.  Of course,
we could have renormalized at a point other than the minimum of the
effective potential (that is, with $\vev/M \ne 1$), and would have
obtained a different expression for the effective potential and a
different relationship between $\lambda$ and $e$, although the physics
would have been the same.  In other words, both descriptions contain
two dimensionless and one dimensional parameters, with only two being
truly independent.  The difference between the two descriptions is
that in the second one dimensional analysis tells us a great deal;
there is a two-parameter family of spontaneously broken theories, but
changing one of the parameters, $\vev$, is completely equivalent to
changing the scale in which masses are measured, so that effectively
there is only a one-parameter family.

Another important point is the singularity in the effective potential 
(and thus in the effective action) at $\pc = 0$.  Because of this
branch point, we cannot expand the effective action in a Taylor 
series about $\pc=0$; the expansion in terms of the $n$-point IPI
Green's functions is not valid.  If it were, we could approximate the 
effective potential near $\pc=0$ (although not everywhere) by the 
sum of the first few Green's functions.  It isn't, and therefore we 
must do an infinite sum even to get an approximation to the effective
potential.

Even if we had taken the theory at face value, as massless scalar 
electrodynamics, there would not have been a simple interpretation of the 
Green's functions in terms of S-matrix elements, because of the 
presence of massless particles.  To calculate any physical process 
we would have had to sum over sets of degenerate states with various
numbers of particles at infinitesimal momenta.

At this point, we can transform the fields as is usually done in 
Higgs models to obtain a massive scalar field and a massive vector
field.  If we define these masses in terms of the values of the 
inverse propagators at zero momentum, we obtain
$$ 
   m^2(S) = \left. {d^2V \over d\pc^2} \right|_{\pc=\vev}
         = {3e^4 \over 8\pi^2} \vev^2 
\eqno(4.31)
$$
and 
$$  
   m^2(V) = e^2 \vev^2
\eqno(4.32)
$$
and thus
$$ 
  {m^2(S) \over  m^2(V) } = {3e^2 \over 8\pi^2} = {3 \over 2\pi} \alpha
    \, .
\eqno(4.33)
$$

It is important to remember that the expression we have obtained for
the effective potential is only valid in the region of classical field
space where $\log(\pc^2/M^2)$ is small; in other words, it is not
valid for very large or very small $\pc$.  Thus, there might be deeper
minima than the one we have found, which lie in regions inaccessible
to our computational methods.  However, the value of the effective
potential at the origin will remain zero, even if higher-order effects
cause this to be a minimum rather than the maximum which the one-loop
approximation indicates.  Since the effective potential is negative at
the minimum which we have found, and since this minimum is in a region
where higher-order effects are expected to be small, we see that the
absolute minimum cannot occur at the origin; spontaneous symmetry
breaking occurs for small values of $e$, at least when $\lambda$ is of
the order of $e^4$.

\medskip
\centerline{4.  Other Gauge Theories}

\medskip

The results for a more complicated (possibly non-Abelian) gauge theory
containing many spin-0 and spin-1/2 particles are, in general,
qualitatively similar to those for scalar electrodynamics.  First, we
note that in these theories the matrices $U(\pc)$, $F(\pc)$ and
$G(\pc)$ have a simple physical interpretation: when evaluated at the
vacuum expectation value of $\phi$, they are the zero-loop
approximations to the mass matrices of the spin-0, spin-1/2, and
spin-1 particles, respectively.  If the vector meson masses are
sufficiently larger than those of the scalar mesons and fermions, the
effects of the scalar and fermion loops will be negligible compared to
those of the vector loops.  (Remember that the contribution of the loop
diagrams is roughly proportional to the fourth power of the relevant
mass.)  In this case, we obtain 
$$
   V =  V_0(\pc)
    +{3e^4 \over 64\pi^2}  {\rm Tr}\,\left\{
          G(\pc)^2 \left[\log{G(\pc)\over M^2} - {25\over 6}
     \right]\right\}
\eqno(4.34)
$$
as the one-loop approximation to the effective potential.  If there 
is a single multiplet of scalar particles whose self-interactions are
described by a single quartic self-coupling, we can absorb $V_0$
into the vector loop by replacing $M$ by $\mu$, where $\mu$, unlike 
$M$, is not arbitrary, since its value determines the strength of the 
scalar self-interaction.  We can then write
$$
    V= {3e^4 \over 64\pi^2} {\rm Tr}\,\left\{
          G(\pc)^2 \left[\log{G(\pc)\over \mu^2} - {1\over 2}
     \right]\right\}  \, .
\eqno(4.35)
$$
This equation is of the same form as Eq.~(4.30); we expect it to have
the same consequences: a minimum in the effective potential, which 
gives rise to spontaneous symmetry breaking and a single relationship
among masses.  Otherwise, we expect theories with our mass renormalization
condition to be similar to the more usual ones with a negative mass
term; by adding one restriction, we have obtained one new result.

To illustrate all this, we consider the Weinberg-Salam model of
leptons,$^8$ modified by requiring that the scalar mass term in the
Lagrangian vanish.  This model is based on an ${\rm SU(2)} \times {\rm
U(1)}$ gauge symmetry; the gauge fields are denoted by ${\vec W}^\mu$ and 
$B^\mu$, with coupling constants $g$ and $g'$, respectively.  There is
a single complex scalar doublet; with no loss of generality, we can 
assume that only the real part of one of its components has a non-zero
vacuum expectation value.  Because the leptons are light, their effects
on the effective potential can be neglected.  We assume that the final
massive scalar meson is light enough that the scalar loop contributions
can also be neglected.  The matric $G(\pc)$ is
$$
   G(\pc) = \left(\matrix{ 
    {1\over 4}g^2 \pc^2  &0 &0 &0 \cr \cr
    0 & {1\over 4}g^2 \pc^2  &0 &0 \cr \cr
    0 & 0 & {1\over 4}g^2 \pc^2 & {1\over 4}g g' \pc^2 \cr \cr
    0 & 0 &  {1\over 4}g g' \pc^2 &  {1\over 4} {g'}^2 \pc^2
      }\right)  \, .
\eqno(4.36)
$$
Its eigenvalues and the corresponding eigenvectors are 
$$ \eqalign{
    {1\over 4}g^2 \pc^2 \, &: \, W_1^\mu \cr
    {1\over 4}g^2 \pc^2 \, &: \, W_2^\mu \cr
    0 \, &: \, A^\mu \equiv {1\over \sqrt{g^2 + {g'}^2}} \,
                 (gB^\mu - g'W^\mu_3)  \cr
   \left({g^2 + {g'}^2 \over 4}\right) \pc^2 \, &: \, Z^\mu \equiv 
          {1\over \sqrt{g^2 + {g'}^2}} \, (g' B^\mu + g'W^\mu_3) \, .}
\eqno(4.37)
$$

The $W_1^\mu$ and $W_2^\mu$ correspond to the charge weak intermediate
vector boson, The $A^\mu$ to the photon, and the $Z^\mu$ to a massive
neutral vector boson.  Identifying the coupling constant of the photon 
as the electric charge, we find
$$
     e^2 = {g^2 {g'}^2 \over g^2 + {g'}^2}  \, .
\eqno(4.38)
$$
Knowing the eigenvalues of $G(\pc)$, we can easily calculate the one-loop
approximation to the effective potential, and see that the minimum 
occurs when $\pc$ is equal to $\mu$.  Thus, the massive vector mesons
have masses
$$
   m^2(W) = {1\over 4} g^2 \mu^2 
\eqno(4.39)
$$
and \
$$
  m^2(Z) = {1\over 4} ( g^2 + {g'}^2 ) \mu^2   \, .
\eqno(4.40)
$$
The mass of the scalar meson is
$$
    m^2(\phi) = \left. {d^2V \over d\pc^2}\right|_{\pc=\mu} 
     = {3 \over 128\pi^2} \left[2g^4 + ( g^2 + {g'}^2 )^2 
  \right] \mu^2  \, .
\eqno(4.41)
$$
We thus obtain the relation
$$
    m^2(\phi) = {3 \over 32\pi^2}\left( 2g^2 m^2(W^\pm) +  ( g^2 + {g'}^2 )m^2(Z)
     \right) \, .
\eqno(4.42)
$$

A similar analysis can be applied to any of the many gauge theories of
weak and electromagnetic interactions which have been recently
proposed.  One must, however, be careful in treating models with heavy
leptons; if the leptons are made massive enough, the fermion loop
contributions will dominate the effective potential and, as we have
seen, will destroy not only the spontaneous symmetry breaking, but
also the vacuum.

\vfill\eject 

\centerline{V. Massive Theories}

\bigskip

So far, we have restricted ourselves to the consideration of theories
in which the Lagrangian does not contain a mass term; we will now
remove that restriction and consider theories in which mass terms
(either positive or negative) are present.  The models we have
considered previously are special cases of this large class of
theories --- special not because they contain massless particles
(they don't necessarily), but because their only dimensional parameter
is the renormalization point, which is arbitrary and which can be
changed without affecting the physics.  While this is significant, it
does not seem to be the sort of property that should characterize the
transition point between normal theories and spontaneously broken
ones.  Therefore, we might expect quantum corrections to invalidate
the conventional wisdom --- perhaps there are theories which are
spontaneously broken even though the Lagrangian contains a positive
mass-squared, or theories which have a negative mass-squared in the
Lagrangian and yet have a symmetric vacuum.  We shall study a few
models to see if we can find any.

First, we derive some computational rules analogous to Eq.~(4.1-11).
We begin by considering the effects of a scalar mass term, 
$-{1\over 2}m^2\phi^2$, where $m^2$, in general, is a matrix
(with either positive or negative eigenvalues).  For the moment we 
consider $U(\pc)$ to not contain any contribution from the mass
term.  We see immediately that the contribution to the effective
potential from the diagrams with one scalar loop is
$$
\eqalign{
   V_{\rm spin-0~loop} &= {\rm Tr}\, {i\over 2} 
       \int{d^4k \over (2\pi)^4} \log \left(1 
      -{U(\pc) \over k^2 -m^2 +i\epsilon} \right)
     \cr    &=  {\rm Tr}\, {i\over 2}
       \int{d^4k \over (2\pi)^4} \left[\log \left(1
      -{ m^2 + U(\pc) \over k^2  +i\epsilon} \right)
         - \log \left(1
      -{ m^2\over k^2  +i\epsilon} \right) \right] \, .
    } 
\eqno(5.1)
$$
If we now consider $U(\pc)$ to include the contribution from the 
mass term, we obtain
$$ \eqalign{
   V_{\rm spin-0~loop} &= {1 \over 64\pi^2} {\rm Tr}\, 
     \left\{\left[2 U(\pc) \Lambda^2 + U(\pc)^2 
     \left(\log {U(\pc)\over \Lambda^2 } -{1 \over 2} \right) \right]
    \right.\cr & \quad \left.
     - \left[2 U(0) \Lambda^2 + U(0)^2
     \left(\log {U(0)\over \Lambda^2 } -{1 \over 2} \right) \right]
     \right\} \, .
}\eqno(5.2)
$$
The second term is constant in classical field space; its only role it to 
assure that the effective potential vanishes when $\pc=0$.  The scalar 
propagator to be used in prototype graphs is even simpler to calculate;
clearly, Eq.~(4.3) remains true if $U(\pc)$ is interpreted to include
the mass term contributions.  The rules for spin-1/2 and spin-1 fields
are just as simple.  $F(\pc)$ and $G(\pc)$ are now understood to 
include contributions from the appropriate mass terms.  The formulas
for the propagators remain unchanged, while the rules for calculating the
one-loop contributions are altered only by the subtraction of the 
value of the loop at $\pc=0$.

After renormalization, however, the expression for the effective
potential becomes more complicated than previously; this is a
consequence of the fact that $U(\pc)$, $F(\pc)$, and $G(\pc)$ are more
complicated functions of $\pc$ than before, which makes the
derivatives of $V(\pc)$ more complicated also.

\medskip
\centerline{1. Massive $\lambda\phi^4$ Theory}

\medskip

We return to our original model, adding a mass term.  The Lagrangian 
is now
$$
  {\cal L} = {1 \over 2} (\partial_\mu \phi)^2 - {1\over 2}m^2\phi^2
      -{\lambda\over 4!}\phi^4 + {\rm counter{\mathhyphen}terms}
\eqno(5.3)
$$
The renormalization conditions we impose are
$$
   \left. {d^2V \over d\phi_c^2} \right|_{\varphi_c =0} =m^2 \, ,
   \eqno(5.4)
$$
$$
    \left. {d^4V \over d\varphi_c^4} \right|_{\varphi_c=M} = \lambda \, ,
\eqno(5.5)
$$
and
$$ 
   Z(M) =1 \, .
\eqno(5.6)
$$
Notice that we have retained the arbitrary non-zero renormalization
point, even though there is no singularity at $\pc=0$ when $m^2 \ne 0$.
The reason should be clear; if we want to consider our previous model
as a special case of a more general class, we should impose the same
renormalization conditions throughout.  (Although $M$ is still arbitrary,
the dependence of the effective potential on $M$ is no longer determined
by dimensional analysis, since the theory now contains a second
dimensional parameter.)

First, we consider the case where both $m^2$ and $\lambda$ are positive.
A straightforward (but agonizing) calculation yields
$$ \eqalign{
     V &= {m^2 \over 2} \pc^2 + {\lambda\over 4!}\pc^4
    + {1\over 64\pi^2} \left(m^2 +{\lambda\pc^2\over 2}\right)^2
    \log\left( 1+ {{1\over 2} \lambda\pc^2\ \over m^2} \right)
    \cr &\quad 
    - {1\over 64\pi^2} 
     \left[ {1\over 2} m^2\lambda\pc^2
      + {\lambda^2\pc^4 \over \left(m^2 + {1\over 2} \lambda M^2\right)^2 }
   \left( {3\over 8}m^4 + {7\over 8}m^2\lambda M^2 
     +{25\over 96}\lambda^2 M^4 \right)\right)]
    \cr &\quad
   -{\lambda^2\pc^4 \over 256\pi^2} 
     \log\left({m^2 + {1\over 2} \lambda M^2\over m^2} \right) \, .
}\eqno(5.7)
$$
The effective potential vanishes at the origin, as it should.  We note
that the first three terms are always non-negative, while the terms in 
square brackets, which are negative, are negligible unless $\lambda$ is 
large (of the order of $64\pi^2$).  Only the last term can make the 
effective potential turn negative, which evidently will happen if 
$ {1\over 2} \lambda M^2$ is much greater than $m^2$.  Of course, this 
will put the minimum at a small value of $\pc/M$ (but with $\lambda\pc^2
> m^2$), which is outside the region of validity of the one-loop
approximation, but we do see our previous results, for $m^2=0$, emerging as
the limit of the massive ones.

Now we let $m^2$ be negative, and write $m^2 = -\mu^2$.  We obtain
$$\eqalign{
    V &= -{\mu^2 \over 2} \pc^2 + {\lambda\over 4!}\pc^4
    + {1\over 64\pi^2} \left(-\mu^2 +{\lambda\pc^2\over 2}\right)^2
     \log\left({|\mu^2 - {1\over 2} \lambda\pc^2| \over \mu^2} \right)
    \cr &\quad
    + {1\over 64\pi^2}
     \left[ {1\over 2} \lambda \mu^2 \pc^2+ 
     {\lambda^2\pc^4 \over \left(\mu^2 - {1\over 2} \lambda M^2\right)^2}
     \left( -{3\over 8}\mu^4 + {7\over 8}\lambda \mu^2M^2
     -{25\over 96}\lambda^2 M^4 \right)\right)]
    \cr &\quad
   -{\lambda^2\pc^4 \over 256\pi^2}
     \log\left({|\mu^2 - {1\over 2} \lambda M^2|\over \mu^2} \right)
    +{i \over 64\pi} \left[\left({\lambda\pc^2\over 2} -\mu^2 \right)^2
    \theta\left(\mu^2 - {\lambda\pc^2\over 2} \right) - \mu^4 
      \right] \, .
}\eqno(5.8)
$$
Postponing for a moment the discussion of the imaginary part, let us 
investigate the behavior of the real part of the effective potential
as $\mu^2$ is varied while ${1\over2} \lambda M^2$ is held fixed.
There is a singularity at $\mu^2 = {1\over2} \lambda M^2$, but the
one-loop approximation clearly fails near this point, since additional 
loops will bring in additional factors of 
$\log\left({ \mu^2 - {1\over 2} \lambda \pc^2 \over
\mu^2 - {1\over 2} \lambda M^2} \right)$; we will therefore assume that this
represents a failure of computational method rather than a real physical 
effect.  Away from this region, the terms in square brackets are small
(for small $\lambda$), and may be neglected.  Let us now consider the
two limiting cases:

1)  $\mu^2 > {1\over 2} \lambda M^2$ : In this region, both of the
logarithmic terms are small; the effective potential is dominated 
by the zero-loop terms, and has a minimum at $\pc \approx \sqrt{6 \mu^2
\over \lambda}$, in agreement with ancient lore.

2) $\mu^2 \ll {1\over 2} \lambda M^2$ : In this region, the logarithmic
factors dominate, and the position of the minimum is given roughly by
Eq.~(3.17).

We now see how the transition from the 
normal mode to the 
spontaneously broken mode occurs:  With ${1\over 2} \lambda M^2$ held
fixed, we let $m^2$ decrease from a large positive value.  The theory is
normal unit $m^2$ becomes much smaller than ${1\over 2} \lambda M^2$
(roughly ${1\over 2} \lambda M^2\exp\left(- {8\pi^2 \over 3 \lambda}
\right)$).  At this point a minimum develops, but it is in a region of 
small $\pc$, where the one-loop approximation is not reliable.  As
$m^2$ decreases through zero and becomes negative, the minimum begins to 
move outward toward the region where the approximation is valid.  As
$-m^2$ become larger than ${1\over 2} \lambda M^2$, the minimum 
reaches its limiting position,  $\sqrt{- 6m^2
\over \lambda}$.  Of course, we cannot be confident about these 
predictions for the regions of large or small $\pc/M$, but we 
do obtain one important result in which we can have confidence: 
If $|m^2|$ is greater than ${1\over 2} \lambda M^2$, the loop 
diagrams give a small correction, and the classical approximation
predicts the correct qualitative behavior.  

Now we must return to the imaginary part of the effective potential,
which arises because the argument of the logarithms in the loop
integral becomes negative for certain values of $\pc$.  At first one
may be inclined to object that the action, and thus the potential,
must be real, and that we must have made some grave error.  It is true
that the quantized action, which is a functional of the quantized
fields, must be a Hermitian operator, but the effective action, which
is a functional of the classical fields, clearly must be complex.
This becomes obvious if one realizes that the effective action is the
generating functional of the IPI Green's functions, which must be
complex for at least some values of the external momenta.  Still, the
effective potential will normally remain real, since the imaginary
part corresponds to on-shell intermediate states, which normally
cannot be produced if all the external momenta vanish.  However, if
the theory contains a particle with imaginary mass, on-shell
intermediate states can be produced, even though the external momenta
vanish.  Of course, the theory doesn't really contain any particles
with imaginary mass, but we are doing perturbation theory as if it
did, and the Green's functions only have meaning in the context of a
particular perturbation theory; if we redefine the fields so as to
eliminate the appearance of a negative mass-squared, the Green's
functions are correspondingly redefined, and have no simple relation
to the previously defined ones.  (Since we are assuming that the
potential is unchanged by a redefinition of the fields, we must
conclude that the imaginary part vanishes when calculated to all
orders; this does not forbid its presence to any finite order.)

Having convinced ourselves that the imaginary part of the effective
potential is not nonsense, we must see how matters are altered by its
presence.  Since the quantized action, and thus the counter-terms,
must be real, we can only renormalize the real part of the effective
action; that is, we require that
$$
   \left. {d^2({\rm Re\,}V) \over d\phi_c^2} \right|_{\varphi_c =0} 
    = -\mu^2
   \eqno(5.9)
$$
and 
$$ 
    \left. {d^4({\rm Re\,}V) \over d\phi_c^4} \right|_{\varphi_c =M}
    = \lambda   \, .
\eqno(5.10)
$$
(This was done in obtaining Eq.~(5.8).)  We note that the imaginary
part is finite, even without being renormalized.  Also, we looked for
spontaneous symmetry breaking by searching for the minimum of the real
part of the effective potential, but this does not matter; for the
imaginary part has two terms, of which one is independent of $\pc$,
and thus physically meaningless, while the other vanishes unless $\pc
< \sqrt{2\mu^2 \over
\lambda}$, which is always below the neighborhood of the minimum.

\noindent So far, we have been assuming that $\lambda$ is positive; what happens
when $\lambda$ is negative?  In this case, the zero-loop approximation
to the effective potential has no lower bound.  It is possible that
the contributions from diagrams with one or more loops might make the
effective potential turn upward at large $\pc$ (as the one-loop terms
appear to do), but our approximation cannot determine this.  Later we
shall see how to improve our approximation, and will find that this
does not happen.  In any case, there is no minimum within the region
of validity of our approximation, no matter what the sign of $m^2$.

\medskip
\centerline{2. Massive Scalar Electrodynamics}

\medskip

Next we turn to massive scalar electrodynamics, described by the 
Lagrangian
$$ \eqalign{
  {\cal L} &= {1 \over 4} F^{\mu\nu} F_{\mu\nu}
     + {1\over 2} \left( \partial_\mu \phi_1 - e A_\mu \phi_2\right)^2
     + {1\over 2} \left( \partial_\mu \phi_2 + e A_\mu \phi_1\right)^2
    \cr &\quad
    -{1\over 2}m^2 (\phi_1^2 +\phi_2^2)
   - {\lambda\over 4!} (\phi_1^2 +\phi_2^2)^2
    + {\rm counter{\mathhyphen}terms}  \, .
}\eqno(5.11)
$$ If $\lambda$ is much greater than $e^4$, we expect that the
contributions to the effective potential from scalar loop diagrams
will dominate those from photon loop diagrams and that the effective
potential will behave qualitatively like that of the $\lambda \phi^4$
theory.  Therefore, let us assume that $\lambda$ is of the order of
$e^4$, or smaller.  In the one-loop approximation, we should then
neglect the scalar loop diagrams, since they are of the same order of
magnitude as the diagrams with two photon loops.  We obtain for the
effective potential
$$
  V = {1 \over 2}m^2\pc^2 +  {\lambda\over 4!} \pc^4
    +  {3 e^4\over 64\pi^2} \pc^4
          \left(\log{\pc^2 \over M^2} -{25\over 6} \right) \, .
\eqno(5.12)
$$
Here $M$ is, as usual, the value of $\pc$ at which the renormalizations
are done.  There is no restriction on the sign of either $m^2$ or 
$\lambda$.

Now we define a quantity $\mu$, with the dimensions of mass, by
$$
   {\lambda \over 4!} = {3e^4 \over 64\pi^2} \left( \log{M^2 \over \mu^2}
       + {11\over 3} \right) \, .
\eqno(5.13)
$$
We then obtain
$$
   V = {1 \over 2}m^2\pc^2 + {3 e^4\over 64\pi^2} \pc^4
        \left(\log{\pc^2 \over \mu^2} -{1\over 2} \right) \, .
\eqno(5.14)
$$
Note that $\mu$, unlike $M$, is not arbitrary; Eq.~(5.14) does not
contain any redundant parameters.  We see that $\mu$ is the position
of the minimum if $m^2$ vanishes.

To study the behavior of the effective potential, we will need the 
following equations:
$$
   {dV \over d\pc} = \pc \left( m^2 + {3e^4 \mu^2\over 16\pi^2}
      {\pc^2\over \mu^2} \log{\pc^2\over \mu^2} \right) \, ,
\eqno(5.15)
$$
$$
   {d^4V \over d\pc^4} = {9 e^4 \over 8\pi^2} \left( 
       \log{\pc^2\over \mu^2} +{11\over 3}\right) \, .
\eqno(5.16)
$$
It will also be useful to consider the behavior of the function
$f(x) = x \log(x)$, shown in Fig.~11. 

We begin by considering the case of positive $m^2$.  We see that 
there is a minimum at the origin, but that if $m^2$ is
sufficiently small, the effective potential will have a second
minimum arising from the logarithmic term; the situation will be 
as in either Fig.~12a or Fig.~12b.  To determine matters more 
quantitatively, we consider Eq.~(5.15), and see that the condition
for an extremum away from the origin is 
$$
   m^2 = -{3e^4 \mu^2\over 16\pi^2} \,\,  f\left({\pc^2\over \mu^2}\right)
  \, .
\eqno(5.17)
$$

If $m^2 > {3e^4 \mu^2\over 16\pi^2} e^{-1}$, Eq.~(5.17) cannot 
be satisfied; the effective potential behaves as in Fig.~12c.  If
$0< m^2 < {3e^4 \mu^2\over 16\pi^2} e^{-1}$, there will be two 
solutions, one for $\pc^2/\mu^2 < e^{-1}$ and one for 
$\pc^2/\mu^2 > e^{-1}$.  Clearly the former corresponds to a maximum
and the latter to a minimum.  We must now determine whether this is
an absolute minimum.  The effective potential vanishes at the origin;
From Eqs.~(5.14) and (5.17) we see that at $\pc^2/\mu^2 = \beta$
it is
$$
   V(\beta\mu^2) = {\beta\mu^4 \over 4} 
    \left(m^2 - {3e^4 \mu^2\over 32\pi^2} \beta \right) \, .
\eqno(5.18)
$$
The condition for an absolute minimum away from the origin is 
therefore
$$
   \beta > {32\pi^2 m^2 \over 3e^4 \mu^2} \, ,
\eqno(5.19)
$$
where $\beta$ satisfies
$$ 
   \beta \log \beta = -{16\pi^2 m^2 \over 3e^4 \mu^2} \, .
\eqno(5.20)
$$
After a little algebra, we see that this condition is satisfied
if $\beta > e^{-1/2}$.  Thus, there three regions:

1)  $0<m^2 < {3e^4 \mu^2\over 32\pi^2}e^{-1/2}$ : The effective potential
behaves as in Fig.~12a; spontaneous symmetry breaking occurs.

2) ${3e^4 \mu^2\over 32\pi^2}e^{-1/2} <m^2 
< {3e^4 \mu^2\over 16\pi^2}e^{-1} $ : The effective potential
behaves as in Fig.~12b; the vacuum is symmetric.

3) $m^2 > {3e^4 \mu^2\over 16\pi^2}e^{-1} $ : The effective potential
behaves as in Fig.~12c; again there is no spontaneous symmetry
breaking.

The results for region (1) seem in contradiction with the conventional 
wisdom --- they predict spontaneous symmetry breaking for a theory 
with a positive mass-squared.  To show that there is no contradiction,
we calculate $\lambda$, using Eq.~(5.16):
$$
   \lambda = \left.{d^4V \over d\pc^4}\right|_{\pc=M} =
    {9e^4 \over 8\pi^2} \left(\log {M^2 \over \mu^2} +{11\over 3}
      \right) \, .
\eqno(5.21)
$$
Normally, we would choose $M$ to be of the order of magnitude of $m$,
since $m$ is the only dimensional parameter in the initial 
formulation of the theory.  (It is true that there is another dimensional
parameter, the position of the minimum of the effective potential, which 
is much greater than $m$, but this is not originally evident.)  If we
choose $M \approx m$, and let $m^2$ be in region (1), we find that 
$\lambda$ is negative (unless $e^4$ is absurdly large).  This is not
the theory of which the conventional wisdom speaks; instead, it is one
which is usually discarded as not having a lower bound on the effective
potential.  However, we see that the effect of the quantum corrections is
to make the effective potential turn upward at large $\pc$, and thus
remedy the situation.  To obtain the more usual theory, with a positive
$\lambda$, $m^2$ must be in either region (2) or region (3), where the 
vacuum is, in fact, symmetric.

Now we turn to the case of negative $m^2$.  There is a maximum at the
origin, and a minimum determined by 
$$
   f\left({\pc^2\over \mu^2}\right) = {\pc^2\over \mu^2}
     \log {\pc^2\over \mu^2} = -{16\pi^2 m^4 \over 3 e^4 \mu^2} 
   = {16\pi^2 \over 3e^4\mu^2} |m^2| \, .
\eqno(5.22)
$$
We see from Fig.~11 that this equation has only one solution.
For $|m^2| \ll \mu^2$, this solution is at $\pc \approx \mu$, which is
in agreement with our results for $m^2=0$.  In general, the minimum
occurs when
$$
   |m^2| = {3e^4 \over 16\pi^2} \pc^2 \log {\pc^2\over \mu^2}  \, .
\eqno(5.23)
$$

To determine the mass of the scalar particle, we calculate
$$\eqalign{
   m^2(\phi) &= \left.{d^2V \over d\pc^2} \right|_{\pc=\vev}
      = -m^2 +{3e^4 \over 16\pi^2} \vev^2 
     \left(3 \log {\vev^2 \over \mu^2} + 2\right) 
    \cr &= 2|m^2| + {3e^4 \over 8\pi^2} \vev^2 \, .
}\eqno(5.24)
$$
The first term is the usual (zero-loop) result for modes of this 
type, while the second term is just our result for the case of
$m^2=0$.  As $\lambda$ (defined at the minimum of the effective
potential) is increased above $e^4$, we find that 
$$
   \vev^2 \approx {6|m^2| \over \lambda}
\eqno(5.25)
$$
and 
$$
   m^2(\phi) = 2|m^2| \left(1+ {9e^4 \over 8 \pi^2 \lambda} 
        \right) \, .
\eqno(5.26)
$$
For $\lambda$ much greater than $e^4$, the second term on the 
right hand side of Eq.~(5.26) is negligible, and the classical 
result holds. 

\medskip
\centerline{3. Electrodynamics with Massive Photons}

\medskip

A model which is useful for understanding the physical principals
behind our results is the theory of a massive charged scalar meson
coupled to a massive photon.$^9$  If we assume that the scalar 
quartic coupling constant, $\lambda$, is of the order of $e^4$, 
we need only consider diagrams with photon loops in the one-loop
approximation.  The expression we obtain for the effective 
potential is 
$$ \eqalign{
     V  &= {\lambda\over 4!} \pc^4 +{3 \over 64\pi^2}  \left\{
    (m^2 + e^2\pc^2) \log\left(1 + {e^2\pc^2 \over m^2} \right)
   - e^2\pc^2 m^2 
   \right. \cr &\quad \left.
   - {e^4 \pc^4 \over (m^2 + e^2 M^2)^2}
    \left({3\over 2}m^4 +7e^2m^2M^2 +{25\over 6}e^4M^4 \right)
 \right. \cr &\quad \left.   
  + e^4 \pc^4 \log \left({m^2 \over m^2 +e^2M^2 }\right) \right\} \, ,
}\eqno(5.27)
$$
where $M$ is the value of $\pc$ at which $\lambda$ is defined
and $m$ is the mass of the photon.  We see that if $m^2$ is much 
larger than $e^2M^2$, the effective potential does not possess a 
minimum away from the origin.  In other words, the coupling to 
the photon can produce spontaneous symmetry breaking only if the
photon is sufficiently light.  But this is equivalent to saying
that spontaneous symmetry breaking occurs only if the range of the 
electromagnetic interaction is sufficiently long.  This agrees with
physical intuition, since spontaneous symmetry breaking is 
essentially a correlation of the field through all of space, which 
should require long-range forces.  It is true that long-range
forces are not evident in the usual examples of the Goldstone
phenomenon, where spontaneous symmetry breaking is induces by
a negative mass-squared term in the Lagrangian, but this is 
such an unnatural term that we really don't have any physical 
interpretation for it until we redefine the fields.  (And when
we do redefine the fields, we obtain massless particles, and 
thus long-range forces.)

\medskip
\centerline{4. A Model with Pseudo-Goldstone Bosons}

\medskip

Finally we consider a model in which the quantum corrections must be
considered in order to completely determine the character of the
vacuum, even though the Lagrangian contains a negative mass-squared
term.  The model is based on an SU(3) gauge symmetry; in addition to the
gauge vector mesons, it has an octet of spinless mesons, $\phi$, which 
we write as a $3\times 3$ traceless Hermitian matrix.  If we impose the
requirement that the Lagrangian be invariant under the transformation
$\phi \rightarrow -\phi$, the most general form for the 
non-derivative scalar meson terms in the Lagrangian is
$$ 
  {\cal L}_{\rm scalar} = -\mu^2 {\rm Tr \,}\phi^2
    + a\left({\rm Tr \,}\phi^2\right)^2 
    + b {\rm Tr \,}\phi^4 \, .
\eqno(5.28)
$$
However, we note that for three-dimensional traceless Hermitian
matrices
$$
    {\rm Tr \,}\phi^4 = {1\over 2}\left({\rm Tr \,}\phi^2\right)^2 \, .
\eqno(5.29)
$$
Thus the zero-loop approximation so the effective potential is of 
the form
$$
   V_0 = -\mu^2 {\rm Tr \,}\phi^2 
   +  \lambda \left({\rm Tr \,}\phi^2\right)^2  \, .
\eqno(5.30)
$$
We note that this expression involves only ${\rm Tr}(\phi^2)$, which
is invariant under the transformations of SO(8); minimizing it will 
determine only ${\rm Tr}(\phi^2)$, but not ${\rm Det}(\phi)$, which is
an invariant under SU(3) transformations but not under SO(8)
transformations.  Yet, because of the presence of the vector mesons 
and their couplings to the scalar mesons, the Lagrangian is invariant
only under SU(3), and not under SO(8).  Therefore, we must 
calculate the one-loop contributions to the effective potential, 
where the effects of the vector mesons first appear, in order to 
completely determine the nature of the spontaneous symmetry 
breaking.  

The SO(8) invariance of the zero-loop effective potential has 
another important consequence:  The spontaneous 
breaking of SO(8) symmetry gives rise to a single massive scalar and
seven massless Goldstone bosons; we will obtain the same result in the
zero-loop approximation to our model, since to that order our model is the 
same as an SO(8) symmetric one.  However, when SU(3) is spontaneously 
broken, either ${\rm SU(2)}\times{\rm U(1)}$ or ${\rm U(1)}\times{\rm U(1)}$
remains as an unbroken subgroup; only four or six of the Goldstone 
bosons can be eaten by the vector meson through the Higgs-Kibble 
mechanism.  The remaining massless bosons acquire a mass when loop 
effects are included; they are only pseudo-Goldstone bosons.$^{10}$

The most general form for the vacuum expectation value of $\phi$ can be
written as
$$
  \vev = \left(\matrix{ a & 0 & 0 \cr 0 & b & 0 \cr
            0& 0& c} \right) \, , \quad a+b+c=0 \, .
\eqno(5.31)
$$
If $a=b$ (or $a=c$, etc.), there will be an unbroken 
${\rm SU(2)}\times{\rm U(1)}$ subgroup; otherwise the unbroken 
subgroup will be ${\rm U(1)}\times{\rm U(1)}$.  We assume that 
the vector meson masses are enough larger than those of the scalar 
mesons that only the contribution to the effective potential
from the vector meson loops is important.  The matrix $G(\vev)$
is calculated to be 
$$
   G(\vev) =2g^2 \left(\matrix{ 
     (b-a)^2 & 0 & 0 & \cdot & \cdot & & & 0 \cr
      0 & (b-a)^2 &&&&&&\cr
      0 && 0  &&&&& \cr
    \cdot  &&& (a-c)^2 &&&& \cr
     \cdot &&&&  (a-c)^2 &&& \cr
      \cdot &&&&& (b-c)^2 && \cr
         &&&&& & (b-c)^2 & \cr
       0 &&&&&&& 0 }\right) \, .
\eqno(5.32)
$$
(Note that there will be either four or two massless vector mesons
depending on whether of not $a=b$ (or $a=c$, etc.), in agreement
with our above remarks.)  We can write the one-loop contribution
to the potential as 
$$ \eqalign{
      V_{\rm loop} &= {3g^4 \over 8\pi^2} \left\{
   (a-b)^4 \left(  \log {(a-b)^2 \over \mu^2} -{1\over 2}\right)
  + (b-c)^4 \left(  \log {(b-c)^2 \over \mu^2} -{1\over 2}\right)
      \right. \cr &\quad
  + (c-a)^4 \left(  \log {(c-a)^2 \over \mu^2} -{1\over 2}\right)
       \, ,
}\eqno(5.33)
$$
where $\mu$ is a mass determined by the vacuum expectation value
of ${\rm Tr}(\phi^2)$.  Since we are interested in the relative
magnitude of $a$, $b$, and $c$, but not in their absolute
magnitude, we minimize the one-loop contributions, leaving $\mu$
fixed.  (We ignore the zero-loop terms, since they do not depend
on the relative magnitude of $a$, $b$, and $c$.)  One can 
readily believe, and somewhat less readily prove, that there is
a minimum when
$$
   a=b={\mu \over 3} \, \qquad  c = -{2\mu \over 3} \, ,
\eqno(5.34)
$$
and that this minimum is unique (aside from permutations of
$a$, $b$, and $c$.)  Thus, there will be an unbroken
${\rm SU(2)}\times{\rm U(1)}$ subgroup.

Next we turn our attention to the spinless mesons.  We identify
the physical meson fields with the perturbations about the minimum;
we denote these by the usual notation for the pseudoscalar 
octet.  These are related to diagonal perturbations by
$$ 
   \eqalign{ 
  a &= {\mu\over 3} + {1 \over \sqrt{2}} \pi^0 
            + {1 \over \sqrt{6}}\eta  \, , \cr
  b &= {\mu\over 3} - {1 \over \sqrt{2}} \pi^0
            + {1 \over \sqrt{6}}\eta  \, , \cr
  c &= -{2\mu\over 3}  -{2 \over \sqrt{6}}\eta  \, .
} \eqno(5.35)
$$
Substituting this into Eq.~(5.33), we find
$$   
   \Delta m^2(\pi^0) = {3g^4 \mu^2 \over \pi^2 }
\eqno(5.36)
$$
and
$$
   \Delta m^2(\eta) = {9g^4 \mu^2 \over \pi^2 } \, ,
\eqno(5.37)
$$
where $\Delta m^2$ indicates the contribution of the one-loop
terms to the scalar meson masses.  Since isospin is an unbroken
symmetry, the results for the $\pi^+$ and $\pi^-$ must be the 
same as that for the $\pi^0$.  Furthermore, since there are only
four massive scalars, the $K$'s must all be Goldstone bosons, 
which are eaten by the four vector mesons which become massive.
Finally, we note that the $\eta$ is clearly the meson which
acquired a mass in zeroth order; the pions are the pseudo-Goldstone
bosons, and their entire mass is given by Eq.~(5.36).

Now we use Eqs.~(5.32) and (5.34) to obtain the vector meson 
masses; we find that the four massive vector mesons have a 
mass given by 
$$
    m^2(V) = 2g^2 \mu^2 \, .
\eqno(5.38)
$$
We thus have the relationship 
$$
   {m^2(\pi) \over m^2(V)} = {3g^2 \over 2 \pi^2}  \, .
\eqno(5.39)
$$

\vfill \eject

\centerline{VI. The Renormalization Group}

\bigskip

In this chapter we show that for the class of theories in which the
Lagrangian does not contain any masses (or any other dimensional
parameters) it is possible to improve the one-loop approximation to
the effective potential by the use of renormalization group
methods.$^{11}$  Recall that the validity of the one-loop approximation
required that both the coupling constants and the logarithms (e.g., 
$\log(\pc^2/M^2)$) be small.  However, these are not two completely
independent conditions; $M$ is arbitrary in that a variation of $M$
can be compensated for by a suitable variation of other quantities
in the theory, leaving the physics unchanged.  For this to be so,
it is clear that the dependence of the effective action on the 
coupling constants and on $M$ must be intertwined.  The nature of the
interdependence is most easily determined in theories which contain 
no dimensional parameter other than $M$, for in these the dependence on
$M$ is determined by dimensional analysis.

To formulate these considerations more exactly, let us consider a 
theory with $n$ coupling constants $\lambda_i$ and $m$ fields
$\psi_j$.  The $\psi_j$ include all fields, regardless of their
spin.  The statement that a variation of $M$ can be compensated for
by a variation of the $\lambda_i$ and of the normalization of the
$\psi_j$ is
$$
  0 =\left[ M {\partial \over M} + \sum_i \bar\beta_i(\lambda)
   {\partial\over \lambda_i} + \sum_j\bar\gamma_j(\lambda) 
   \int d^4x \, \psi_{cj}(x) {\delta\over \delta\psi_{cj}(x) }
   \right] \Gamma \, ,
\eqno(6.1)
$$
where $\Gamma$ is the effective action.  The $\bar\beta_i$ and the
$\bar\gamma_j$ can be functions of only the $\lambda_i$, since there
are no other dimensionless parameters available.  If we apply this
equation to the expansion of $\Gamma$ in terms of the IPI Green's
functions,  we obtain the usual equation of the renormalization 
group.  However, as we have said before, this is not the best 
expansion for our purposes.  Instead, we work with the Taylor
series expansion of $\Gamma$.  Since each term in this expansion 
is identified by the power of momentum and the number of fields with 
spin, the action of the differential operator of Eq.~(6.1) does not
mix terms, and so Eq.~(6.1) must hold term by term.  Let us
specialize to the case where there is only a single spinless 
field $\phi(x)$, which we have now expanded about $\Phi_c(x) = \pc$, 
and any number of fields with spin.  (The results are unchanged if we 
replace the single spinless field by an SO($n$) vector of spinless
fields.)  A typical term in our expansion of $\Gamma$ will contain
$n_j$ fields of type $j$ (possibly including $\Phi_c(x)$), some
power of momentum, and a dimensionless function $F(\pc,\lambda_i,M)$.
Analogous to Eq.~(6.1), we will obtain 
$$
   \left[ M {\partial \over M} + \sum_i \bar\beta_i(\lambda)
   {\partial\over \lambda_i} 
    + \sum_j\left(n_j \bar\gamma_j(\lambda) \right)
    +\pc {\partial\over \pc}\right] F(\pc,\lambda_i,M)
\eqno(6.2)
$$
Now we use the fact that $M$ is the only dimensional parameter 
in the theory; dimensional analysis tells us that $F$ can be
a function of only $\pc/M$ and the $\lambda_i$.  If we 
define
$$
   t =\log {\pc \over M} \, ,
\eqno(6.3)
$$
$$   
    \beta_i = {\bar \beta_i \over 1 -\bar\gamma_\varphi}  \, ,
 \eqno(6.4)
$$
and 
$$
    \gamma_j = {\bar\gamma_j \over 1 -\bar\gamma_\varphi} \, ,
\eqno(6.5)
$$
we can rewrite Eq.~6.2) as 
$$
   \left[ -{\partial \over \partial t} \
        + \sum_i \beta_i(\lambda) {\partial\over \lambda_i}
      + \sum_j n_j \gamma_j(\lambda) \right] F(t,\lambda) \, .
\eqno(6.6)
$$
Assuming that we know the $\beta_i$ and the $\gamma_j$, we
can write down the general solution to Eq.~(6.6).$^{12}$  
It is
$$
  F(t,\lambda) = f(\lambda'(t,\lambda))
   \exp\int_0^t dt' \sum_jn_j\gamma_j(\lambda'(t',\lambda))
   \, ,
\eqno(6.7)
$$
where $\lambda'_i(t,\lambda)$ is determined by
$$
    {\partial \lambda'_i(t,\lambda) \over \partial t} 
       = \beta_i(\lambda')
\eqno(6.8)
$$
and 
$$
    \lambda'_i(0,\lambda) = \lambda_i \,
\eqno(6.9)
$$
and $f$ is an arbitrary function of $\lambda'$.

These equations are very pretty, but they involve the $\beta_i$ and
the $\gamma_j$, which we don't know until we have solved the theory
exactly, at which point we don't need the equations.  However, our
perturbation method allows us to obtain approximations for the
$\beta_i(\lambda')$ and the $\gamma_j(\lambda')$ which are valid for
small $\lambda'$: We use the one-loop calculations to obtain
$\beta_i(\lambda'(0,\lambda))$ and $\gamma_j(\lambda'(0,\lambda))$,
and then use Eq.~(6.8) to obtain an expression for
$\lambda'_i(t,\lambda)$ which is valid in that range of $t$ where
$\lambda'$ remains small.  What we have gained in comparison with our
previous method is that $\lambda'$ may remain small even when $t$ is
large, so that our new results will be valid in this region, even
though our old ones were not.  Let us now demonstrate the application
of these methods to some models.

\medskip
\centerline{1. $\lambda \phi^4$ Theory}

\medskip 

Because of its simplicity, we return to our original model.  In this
section we modify it slightly by choosing Eq.~(3.11b) as the coupling
constant renormalization condition.  We can then define a function
$U(t,\lambda)$ by 
$$   
    V(\pc) = {\pc^4 \over 4!} \, U(t,\lambda)  \, .
\eqno(6.10)
$$
Our coupling constant and wave function renormalization conditions
are then
$$
    U(0,\lambda) = \lambda
\eqno(6.11)
$$
and
$$   
   Z(\lambda) = 1  \, .
\eqno(6.12)
$$
 
If we substitute $Z(t,\lambda)$ and $U(t,\lambda)$ for $F(t,\lambda)$
in Eq.~(6.6), and evaluate at $t=0$, we obtain
$$
   - \left.{\partial Z\over \partial t}\right|_{t=0}  
          + 2 \gamma(\lambda) =0
\eqno(6.13)
$$
and
$$
   - \left.{\partial U\over \partial t}\right|_{t=0}
     +\beta(\lambda) + 4 \lambda \gamma(\lambda) =0 \, .
\eqno(6.14)
$$
Furthermore, we can substitute $Z(t,\lambda)$ and $U(t,\lambda)$ for
$F(t,\lambda)$\ in Eq.~(6.7), again evaluate at $t=0$, and obtain
$$
    Z(0,\lambda) = 1 = f_Z(\lambda'(0,\lambda)  = f_Z(\lambda)
\eqno(6.15)
$$
and 
$$
    U(0,\lambda) = \lambda = f_U\lambda(\lambda'(0,\lambda)  
            = f_U(\lambda)  \, .
\eqno(6.16)
$$
Knowing the form of $f_Z$ and $f_U$, we can now write
$$
    Z(t,\lambda) = \exp\left[ 2\int^t_0 dt'\gamma(\lambda'(t',\lambda))
        \right]
\eqno(6.17)
$$
and 
$$  U(t,\lambda) = \lambda'(t,\lambda) \left[Z(t,\lambda)\right]^2  \, .
\eqno(6.18)
$$
Using Eqs.~(6.13), (6.14), (6.17), and (6.18), and one-loop
calculations for $Z(t,\lambda)$ and $U(t,\lambda)$, we obtain
an improved approximation for the effective potential.

Eq.~(6.18) enables us to gain further understanding of the physical 
significance of $\lambda'(t,\lambda)$.  Suppose we change
the renormalization point from $M$ to $M^*$.  We obtain a
rescaled classical field $\Phi_c^*$, given by 
$$
   \Phi_c^* = \left[ Z\left( \log{M^* \over M}, \lambda \right)
    \right]^{1/2} \Phi_c \, ,
\eqno(6.19) 
$$
and a new coupling constant $\lambda^*$, given by 
$$\eqalign{
    \lambda^* &= {4! \over (\pc^*)^4} \left. V\right|_{\pc =M^*}
  \cr        &= \left[ Z\left( \log{M^* \over M}, \lambda \right)
    \right]^{-2} U\left(\log{M^* \over M}, \lambda \right)  \cr
    &= \lambda'\left(\log{M^* \over M}, \lambda \right)
    \, .
}\eqno(6.20)
$$
We now proceed to our calculations.  From our results in Chapter~III,
we obtain the one-loop approximation for $U(t,\lambda)$; it is
$$
   U(t,\lambda) = \lambda + {3\lambda^2 \over 32\pi^2} 
         \log{\pc^2 \over M^2} 
       = \lambda + {3\lambda^2 t\over 16\pi^2} \, .
\eqno(6.21)
$$

The one-loop approximation to $Z(t, \lambda)$ is obtained from the
sum of all one-loop graphs which have two external lines carrying 
momenta $p$ and $-p$, respectively, and all other external momenta
vanishing; we call the sum of these graphs $\Sigma(p^2)$.
$Z(t, \lambda)$ is given by 
$$
   Z(t, \lambda) = {d \over dp^2} \left.\Sigma(p^2)\right|_{p^2=0}
\, .
\eqno(6.22)
$$
Dimensional analysis simplifies matters considerably; before 
renormalization $Z$ must be a dimensionless function of 
$\pc^2/\Lambda^2$, so only the logarithmically divergent terms in 
$Z$ will survive after renormalization.  Thus we need consider
only the logarithmically divergent part of $Z$, which arises
from the quadratically divergent graphs in $\Sigma(p^2)$.  For the
theory we are considering, there is only one such graph, shown in
Fig.~13.  Furthermore, the value of this diagram is clearly 
independent of $p^2$, so the one-loop approximation to 
$Z(t,\lambda)$ vanishes, and in the one-loop approximation we
have
$$
   Z(t,\lambda) =1 \, .
\eqno(6.23)
$$
 
If we substitute Eqs.~(6.21) and (6.23) into Eqs.~(6.13)
and (6.14), we obtain 
$$ 
  \gamma(\lambda) = 0
\eqno(6.24)
$$
and 
$$ 
    \beta(\lambda) = {3 \lambda^2 \over 16\pi^2}  \, .
\eqno(6.25)
$$
Thus we determine $\lambda'$ from 
$$
    {d \lambda' \over dt} = {3 {\lambda'}^2 \over 16\pi^2} 
\eqno(6.26)
$$
together with the boundary condition, Eq.~(6.9), and 
obtain
$$
   \lambda'(\lambda,t) =  {\lambda \over 
                    1 -{3\lambda t \over 16\pi^2}}  \, .
\eqno(6.27)
$$
Therefore,
$$\eqalign{
    V(\pc) &= {\pc^4 \over 4!} U(t,\lambda) = 
                   {\pc^4 \over 4!} \lambda'(\lambda,t)
   \cr &= {1\over 4!} {\lambda \pc^4 \over 
      1- {3\lambda \over 32\pi^2} \log{\pc^2 \over M^2}} 
\, .} 
\eqno(6.28)
$$
How does this compare with our former expression for the
effective potential, Eq.~(3.13) (allowing for the redefinition
of $\lambda$)?  The two expressions agree in the region where
$|\lambda| \ll 1$ and $|\lambda \log(\pc^2/M^2)| \ll 1$, which is the
region where the earlier expression was valid.  However, our new
result is valid as long as we remain in the region where 
$\lambda'$ is small.  First, suppose that $\lambda$ is negative.
In this case, $\lambda'$ remains small as $t$ becomes large and 
positive, so Eq.~(6.28) is valid in the region of large $\pc$;
we see that the effective potential has no lower bound and the
theory does not exist.  Thus, the only physically meaningful case
is that of positive $\lambda$.  In this case, $\lambda'$ remains
small as $t$ becomes large and negative, but as $t$ becomes large and
positive $\lambda'$ has a pole; we cannot continue past the pole.
That is, Eq.~(6.28) is valid for arbitrarily small $\pc$, but
fails for large $\pc$.  It tells us that the effective potential
has a minimum at the origin; Eq.(3.13) predicted a maximum, but was
not a reliable approximation in that region.  Furthermore, 
Eq.~(6.28) does not predict any other minima in its range of
validity; this theory does not appear to display spontaneous
symmetry breaking.

\medskip
\centerline{2. The Scalar SO($n$) Theory}

\medskip

If we repeat this analysis for the SO($n$) model of Chapter IV,
we obtain similar results.  We find that
$$
   \beta = {3 \lambda^2 \over 16\pi^2} \left( 1 + {n-1 \over 9}
       \right)
\eqno(6.29)
$$
and 
$$
   V(\pc) = { {1 \over 4!} \lambda \pc^4 \over 
   1- {3\lambda \over 32\pi^2} \left( 1 + {n-1 \over 9}
       \right)  \log{\pc^2 \over M^2}} 
\, .
\eqno(6.30)
$$
Particularly interesting is the case of very large $n$, where
our result for the effective potential can be approximated by
$$
     V(\pc) = { {1 \over 4!} \lambda \pc^4 \over
   1- {n\lambda \over 96\pi^2}   \log{\pc^2 \over M^2}}
\, .
\eqno(6.31)
$$

There is an alternative method of improving our approximation when $n$
is very large.  We note that the dominant contribution in each order
will be from the diagrams shown in Fig.~14, both because these are
most numerous for large $n$ and because these have the most 
logarithms for a given power of $\lambda$.  Summing these diagrams
is somewhat complicated by the fact that the coupling constant 
at a vertex can be either $\lambda/3$ or $\lambda$.  However, if 
$n$ is large, almost all vertices will have a factor of $\lambda/3$;
if we use this factor for all vertices (this is in the same spirit
as that which motivated Eq.~(6.31)), we obtain the renormalization 
group result, Eq.~(6.31).  In other words, the renormalization 
group approach has enabled us to sum the ``leading logarithms''.

\medskip
\centerline{3. Massless Scalar Electrodynamics}

\medskip

Next we apply these methods to massless scalar electrodynamics.
Our expansion of the effective action is
$$ \eqalign{
  \Gamma =& \int d^4x \left\{ -V(\pc) 
     -{1 \over 4}(\partial^\mu A_c^\nu - \partial^\nu A_c^\mu)^2 H(\pc)
         \right.  \cr
      &  \quad   \left.
   +{1 \over 2}\left[ (\partial_\mu\Phi_{1c})^2 
 +(\partial_\mu\Phi_{2c})^2 \right] Z(\pc)  
        \right.  \cr
      &  \quad   \left.
    + e \left[ -\partial_\mu \Phi_{1c} A^\mu_c \Phi_{2c}
             + \partial_\mu \Phi_{2c} A^\mu_c \Phi_{1c} \right]F(\pc)
    \right.  \cr
      &  \quad  \left.
 +e^2 \left[ \Phi_{1c}^2 A_{\mu c}A^\mu_c 
              + \Phi_{2c}^2 A_{\mu c}A^\mu_c \right] G(\pc) 
          \right.  \cr & \quad \left.
    +\cdots \right\}  \, ,
}
\eqno(6.32)
$$
where the dots represent terms which do not concern us here because
they are not involved with the renormalization procedure.  Our
renormalization conditions are 
$$
    H(M) =1 \, ,
\eqno(6.33)
$$
$$
    Z(M) =1 \, ,
\eqno(6.34)
$$
$$
    F(M) =1 \, ,
\eqno(6.35)
$$
and 
$$
   U(M) = {4! \over \pc^4} \left.V(\pc)\right|_{\pc =M} = \lambda \, .
\eqno(6.36)
$$
(Note that we have modified the definition of the scalar quartic
coupling constant, as in the previous sections.)  We will also 
find that 
$$ 
   G(M) =1  \, .
\eqno(6.37)
$$

If we apply Eq.~(6.6) to $H$, $Z$, $eF$, and $U$ and evaluate at 
$\pc=M$, we obtain
$$
    - {\partial H\over \partial t} + 2 \gamma_A =0 \, ,
\eqno(6.38)
$$
$$
    - {\partial Z\over \partial t} + 2 \gamma_\varphi =0 \, ,
\eqno(6.39)
$$
$$
   - {\partial (eF) \over \partial t} + \beta_e 
        + e\gamma_A + 2e\gamma_\varphi = 0  \, ,
\eqno(6.40)
$$
and
$$
    - {\partial U\over \partial t} + \beta_\lambda 
         + 4 \lambda \gamma_\varphi = 0 \, .
\eqno(6.41)
$$
Once we have calculated $H$, $Z$, $F$, and $U$, we can use
these equations to find the $\beta$'s and the $\gamma$'s,
and then determine $\lambda'$ and $e'$.

Before we calculate the functions $H$, $Z$, $F$, and $G$, we 
should consider the consequences of electromagnetic gauge 
invariance.  First of all, there is the usual result that
$Z_1 = Z_2$, where $Z_1$ and $Z_2$ are the usual rescaling 
factors in the quantized action; they are not the same as any 
of the functions which appear in our expansion of the effective
action.  Because $Z_1 = Z_2$, there is a relationship among
the counter-terms for the scalar wave function renormalization
and the scalar-photon vertices.  The same relationship must hold
among those terms in $Z$, $F$, and $G$ which are logarithmic 
in $\pc$, since these terms were made finite by the counter-terms.
That is, $Z$, $F$, and $G$ are equal, up to terms which are finite
and independent of $\pc$.  The reason that finite differences might 
exist is that these functions are not the same as the functions 
one considers in the usual renormalization procedure; for example,
$Z$ involves a sum of diagrams with various numbers of external 
lines, while the usual scalar self-energy involves only diagrams 
with two external lines.

It is also a usual consequence of the Ward identities that the photon
self-energy vanishes at zero four-momentum and is transverse for 
non-zero four-momentum.  This is still true, but not of immediate use
to us, since we consider the self-energy (which has no external
scalar lines) in conjunction with diagrams with many external
scalar lines.  However, the counter-terms are restricted by this result,
so we should not expect any divergent non-transverse term in the 
photon self-energy.

As we showed previously, we need consider only those diagrams 
which give logarithmically divergent contributions to the 
functions we are calculating.  These diagrams are shown in 
Figs.~15--18.  (Diagrams 18b--d are included because there are
implicit external scalar lines, even though they are not 
shown explicitly on the prototype graph.)   Several of these
diagrams can be neglected:  Diagrams~15b, 15c, 16b, 16c, and 16d
are independent of $p^2$, so they cannot contribute to the wave
function renormalization, while Diagram~17c must vanish if either of 
the external scalar momenta is zero and thus must be at least of
order $p^2$ and finite.  The remaining diagrams must be calculated using
a regulator, such as that of 't~Hooft and Veltman,$^{13}$ which 
preserves the gauge invariance.  Doing this, one obtains
$$
   H = 1 - {e^2 \over 24\pi^2} t 
\eqno(6.42) 
$$
and 
$$
   Z = F= G + 1 + {3 e^2 \over 8\pi^2} t  \, .
\eqno(6.43)
$$
(It is also interesting to note that the contributions from 
Diagrams~18b--d cancel, as might be expected from a naive 
application of the Ward identities.)  We obtain $U$ from our
previous calculation; it is
$$
    U = \lambda + \left( {5\lambda^2 \over 24\pi^2}
        + {9e^4 \over 4\pi^2} \right) t \, .
\eqno(6.44)
$$

We now use Eqs.~(6.38--41) to obtain
$$
    \gamma_A = - {e^2 \over 48\pi^2}  \, ,
\eqno(6.45)
$$
$$
    \gamma_\varphi =  {3 e^2 \over 16\pi^2}  \, ,
\eqno(6.46)
$$
$$ 
    \beta_e = {e^3 \over 48\pi^2}  \, ,
\eqno(6.47)
$$
and 
$$ 
   \beta_\lambda = {1 \over 4\pi^2} \left( {5 \over 6} \lambda^2
       - 3 e^2 \lambda + 9e^4 \right) \, .
\eqno(6.48)
$$
We must then solve
$$
    {de' \over dt} = {3{e'}^3 \over 48\pi^2}
\eqno(6.49)
$$
and 
$$
   {d \lambda' \over dt} = {1 \over 4\pi^2} \left( {5 \over 6} {\lambda'}^2
       - 3 {e'}^2 \lambda' + 9{e'}^4 \right) \, .
\eqno(6.50)
$$
The former equation is easily solved, yielding
$$
    {e'}^2 = { e^2 \over 1 - {e^2 \over 24\pi^2}t}  \, .
\eqno(6.51)
$$
To solve the latter equation, we define $R = \lambda'/{e'}^2$, and 
obtain the equivalent equation
$$
    {e'}^2 {dR \over d({e'}^2) } = 5R^2  - 19R +54 \, .
\eqno(6.52)
$$
This can be easily solved, and we finally obtain
$$
    \lambda' = {{e'}^2 \over 10} \left[ \sqrt{719} \, \tan\left(
          {\sqrt{719} \over 2} \log {e'}^2 +\theta \right)
        + 19 \right] \, ,
\eqno(6.53)
$$
where $\theta$ is chosen so that $\lambda' =  \lambda$ when $e'=e$.

At this point in our consideration of the $\lambda \phi^4$ model we
used Eq.~(6.18) to obtain an expression for the effective potential
which was valid in a larger region than the one we had originally
obtained; can we do this in the present case?  If $t$ becomes large 
and positive (large $\pc$), $e'$ becomes large; it $t$ becomes
large and negative (small $\pc$), $e'$ becomes small but $\lambda'$
becomes large.  Thus, our new expression for the effective potential
is valid only in the region of small $t$, where our previous 
expression was valid.  

If there are minima hiding in regions of very large or very small
$\pc$, we have not found them; we still know only of the minimum we
found originally.  Although we have not found any more minima, we have
not wasted our time in doing these calculations.  Previously, we
required that $\lambda$ be of the order of $e^4$ in order that the
minimum occur; we can now escape this restriction.  Looking at
Eq.~(6.53), we see that as $e'$ is varied over a relatively small
range, the argument of the tangent changes by $2\pi$, so that
$\lambda'$ takes on all possible values.  (Of course, we can only
trust this statement for that part of the variation in which
$\lambda'$ remains small.)  Thus, if $\lambda$ is not of the order of
$e^4$, but is still small, we can change the renormalization point so
that $\lambda$, defined at the new renormalization point, is of the
order of the new $e^4$.  Our original calculations, using the new
$\lambda$, $e$, and $M$, are then valid for small $t$ (that is, $\pc$
near $M$) and we find a minimum in the effective potential.  That is,
for any small values of $\lambda$ and $e$, massless scalar
electrodynamics does not exist.
  
\medskip
\centerline{4. A Massless Yang-Mills Theory}

\medskip

Finally, we consider a non-Abelian gauge theory: the theory of 
an SU(2) gauge particle coupled to a scalar triplet, with the 
Lagrangian
$$ \eqalign{
    {\cal L} =& -{1\over 4} \left( \partial_\mu \vec A_\nu
   - \partial_\nu \vec A_\mu  - g \vec A_\mu \times \vec A_\nu
      \right)^2
   + {1 \over 2} \left( \partial_\mu \vec \phi 
             -g \vec A_\mu \times \vec \phi \right)^2
    - {\lambda \over 4!} \left( \vec \phi \cdot \vec\phi \right)^2
      \cr & \quad  + {\rm counter{\mathhyphen}terms}  \, .   }
\eqno(6.54)
$$
In calculating Feynman diagrams, we must include terms from the 
so-called ``ghost particles''.  In the Landau gauge the couplings
of the fictitious ghost particle are given by the effective
Lagrangian
$$
  {\cal L}_{\rm ghost} = \partial_\mu \bar \chi \partial^\mu \chi
      - g \vec{\bar \chi} \times \vec A_\mu \cdot \partial^\mu
         \vec \chi   \, .
\eqno(6.55)
$$
We note that in this gauge the ghosts do not couple to the 
scalar mesons; as a result, the ghost propagator to be used
in prototype graphs remains
$$
    { i \over k^2 }
\eqno(6.56)
$$
Thus, there will be infrared divergences arising from ghost 
loops, even when $\pc$ is non-vanishing; these infrared 
divergences complicate the renormalization procedure 
considerably.  

The usual way to renormalize a theory such as this would be to choose
some arbitrary point in momentum-space and to do all renormalizations
there, rather than at zero four-momentum.  Furthermore, by seeing how 
the coupling constants and the wave function normalizations 
changed when this point was varied, we could hope to obtain information
about the high or low momentum behavior of the Green's functions.
However, this is not quite what we want; we want to know about the
behavior of the effective potential at high or low field strength, 
which means that we must vary a renormalization point in field
strength-space.  (In the theories we considered previously, both 
methods would have yielded the same $\beta$'s and $\gamma$'s, since
the same diagrams contribute in either method.  For the theory at 
hand this is no longer true; the contribution from ghost 
loop diagrams will change if a momentum-space renormalization 
point is varied, but not if a field strength-space renormalization
point is changed.)

Thus, we must introduce two arbitrary dimensional parameters, one in 
field strength space ($M$) and one in momentum space ($\mu$),  It would
seem that we would have to replace Eq.(6.1) by a pair of more complicated
equations, corresponding to the presence of two arbitrary parameters.
Fortunately, we can avoid this complication, at least for some 
regions of field strength-space.  We note that there are two types 
of one-loop diagrams: those without ghost loops, which are infrared 
convergent, and those with ghost loops, which are infrared divergent but
independent of $\pc$.  Now suppose that we choose $gM$ to be much 
larger than $\mu$, and that we consider field strengths such that
$g\pc$ is much larger than $\mu$.  Since the contribution of the 
first type of diagram is analytic at $\mu=0$, we can approximate it
by its value at that point.  Since the contribution of the second 
type is independent of $\pc$, it disappears from the effective 
action once we subtract at $\pc=M$.

Actually, the constraints imposed by the Ward identities, which must 
be preserved in order that the theory be renormalizable, add a 
further complication.  To illustrate this, consider the wave 
function renormalization of the vector meson.  We denote by 
$\Pi^{\mu\nu}$ the sum of all graphs with two external vector mesons,
with momenta $p$ and $-p$, and any number of scalars with 
vanishing momenta.  We can write
$$
   \Pi^{\mu\nu} = \left( g^{\mu\nu} - {p^\mu p^\nu \over p^2}
        \right) A(p^2, \pc) + {p^\mu p^\nu \over p^2}B(p^2, \pc)
      \, .
\eqno(6.57)
$$
It is tempting to assume that the Ward identities require $B$ to
vanish, but this is not so because we are considering graphs with
external scalars, and not just vacuum polarization graphs.  However,
the Ward identities do require that the counter-terms contribute to
$A$ only; $B$ must be cutoff independent without renormalization.
It is, being of the order of 
$$
     g^2 \log \left({g^2 \pc^2 \over \mu^2} \right) \, .
\eqno(6.58)
$$
$B$ is non-vanishing because the ghost and the non-ghost diagrams
are not separately transverse, although their logarithmically 
divergent longitudinal parts cancel.  A similar effect will arise
in the renormalization of the $A^\mu A^\nu A^\lambda$ and 
$A^\mu A^\nu A^\lambda A^\tau$ vertices; because of the restrictions
on the counter-terms, there will be terms in the effective action 
containing $\log(g^2M^2/\mu^2)$.  Although we can avoid these terms 
in our one-loop renormalization group analysis, they will 
arise in diagrams with two or more loops.  Since we want to be 
able to neglect two-loop graphs, we must require that 
$$
     g^2 \log \left({g^2 M^2 \over \mu^2} \right)
             \ll 1
\eqno(6.59)
$$
and we must stay in a region of field strength-space where
$$  
  g^2 \log \left({g^2 \pc^2 \over \mu^2} \right) 
       \ll 1  \, .
\eqno(6.60)
$$
If we remain within these restrictions, $\mu$ will have a negligible
effect on the renormalization group analysis.

However, we now find that there is an ambiguity in the definition of
$g$.  If we had followed the usual procedure of renormalizing in
momentum-space, we would have obtained the same results whether we
defined $g$ in terms of the $A^\mu A^\nu A^\lambda$ vertex or in terms
of the $\phi \phi A^\mu$ vertex.  Using our procedure, the two
definitions give different results, since a ghost loop contributes to
the former but not to the latter.  The difference is due to having
pushed different amounts of $\log (g^2M^2 / \mu^2)$ into the higher
loop contributions; therefore, if we are to be able to consistently
neglect two-loop graphs, the ambiguity must be small.  We will find
that it is, and that different choices of the renormalization
conditions (even renormalizing in momentum-space) lead to only small
quantitative, but not qualitative, changes.

Let us choose to define $g$ in terms of the $\phi \phi A^\mu$ vertex.
We then have the same renormalization conditions as in scalar
electrodynamics, namely Eqs.~(6.33--36), where the functions are
defined by the obvious generalization of Eq.~(6.32).  The one-loop
diagrams contributing to $H$, $Z$, and $F$ are shown in Figs.~19--21.
Ignoring finite terms of the order of $\mu^2/g^2 M^2$, we obtain
$$
    H = 1 + {23\over 6} {g^2 \over 16\pi^2} \log{\pc^2 \over M^2}
      \, ,
\eqno(6.61)
$$
$$
    Z = 1 +  6 {g^2 \over 16\pi^2} \log{\pc^2 \over M^2}
      \, ,
\eqno(6.62)
$$
and 
$$
    F = 1 + {9\over 2} {g^2 \over 16\pi^2} \log{\pc^2 \over M^2}
      \, .
\eqno(6.63)
$$
From a calculation of the effective potential, we obtain
$$
    U = \lambda + \left(36g^4 +{11\over 6} \lambda^2 \right)
     {1 \over 16\pi^2} \log{\pc^2 \over M^2} \, .
\eqno(6.64)
$$
Following our previous analysis, we obtain from Eqs.~(6.62) and
(6.64)
$$
    \gamma_\varphi = 6 \, {g^2 \over 16\pi^2}
\eqno(6.65)
$$
and 
$$
    \beta_\lambda = {1 \over 16\pi^2} 
         \left(72g^4   -24 \lambda g^2 
    +{11\over 3} \lambda^2 \right)   \, .
\eqno(6.66)
$$
These two results are unaffected by the ambiguity in the definition
of $g$.  We now use Eqs.~(6.61), (6.63), and (6.65) to obtain
$$
      \gamma_A = {23 \over 6} \, {g^2 \over 16\pi^2}
\eqno(6.67) 
$$
and 
$$  
     \beta_g = - {41\over 6} \, {g^2 \over 16\pi^2} \, .
\eqno(6.68)
$$
These results are affected by a change in the renormalization conditions;
however, the effect is slight (e.g., the $-41/6$ in $\beta_g$
is replaced by $-7$ if renormalization is done in momentum-space).

Proceeding as before, we obtain
$$
    {g'}^2 = { g^2 \over 1 +{41\over 3} {g^2 t \over 16\pi^2}}
\eqno(6.69)
$$
and 
$$
    \lambda' = - {41 \over 22}{g'}^2 \left\{ {41\over \sqrt{8543}}
        \, \tan\left(
          {\sqrt{8543} \over 82} \log {g'}^2 +\theta \right)
         - {31 \over 41} \right] \, .
\eqno(6.70)
$$
These results are reminiscent of the results for scalar electrodynamics,
Eqs.~(6.51) and (6.53).  However, we see that $g'$ is small
when $t$ is large and negative, which is just the opposite of the
manner in which $e'$ behaves.  As before, $\lambda'$ becomes 
large in either case, so the range of validity of the one-loop
approximation to the effective potential cannot be extended; 
however, we can remove any restriction on the relative magnitude
of $\lambda$ and $g$.

\vfill\eject 

\centerline{VII. Conclusions}

\bigskip

Our most important result is the discovery that spontaneous symmetry 
breaking can be induced by radiative corrections in a theory which 
at first glance appears to have a symmetric vacuum; in particular, this
occurs in massless gauge theories.  It is true that we have based this
conclusion on perturbation theory calculations, but there does not
seem to be any obvious reason why it should be any less plausible than
any other conclusion based on perturbative calculations in 
quantum field theory, at least for small coupling constants.

Although we have obtained interesting results in some of the theories
we have considered, it does not appear that they will enable us to
formulate a renormalizable theory of the weak and electromagnetic
interactions that will be any less ugly than the plethora of existing
models.  Rather, it seems profitable to abstract a physical principle
from our results: When the infrared divergences in a massless theory
become unbearable, nature avoids them by the introduction of an
asymmetric vacuum.  If we are willing to accept this speculation, we
may further speculate that this principle remains true even when the
theory does not contain any fundamental scalar fields, that is, when
some compound field develops the vacuum expectation value that induces
the spontaneous symmetry breaking.  We may then hope that when a
satisfactory renormalizable model of the weak and electromagnetic
interactions is found, it will be possible to show that it is
equivalent to a theory containing neither negative mass-squared terms
nor Higgs scalars.  The latter theory, like the ones we have
considered, would have relationships among its coupling constants;
pushing our optimism to a probably absurd level, we might hope that it
contained only one parameter, the fine structure constant, which would
then be determined.

Thus it seems clear that it is desirable to apply our methods to
theories without scalar particles to see if any can be found in which
radiative corrections induce spontaneous symmetry breaking.  As
mentioned in Chapter~II, our formalism can be readily applied to the
study of such theories.  Unfortunately, this search has so far been
fruitless.

\vfill\eject 

\centerline{Appendix}
\centerline{Connection with a Theorem of Georgi and Glashow}

\bigskip

There is a theorem due to Georgi and Glashow$^{14}$ which states that
if the Lagrangian contains an unbroken subgroup of its symmetry group,
the relationships arising from the presence of the unbroken symmetry
remain true even when the parameters in the Lagrangian are changed by
a small amount (for example, by radiative corrections), unless the
theory contains a massless particle which is not a Goldstone boson.
Our results for massive scalar electrodynamics appear to contradict
this result; we found that, for a sufficiently small mass, the
one-loop corrections to the effective potential could induce
spontaneous symmetry breaking, even though the Lagrangian contained an
unbroken symmetry.  The resolution of this conflict lies in the
requirement that the variation of the parameters be small; the
one-loop corrections are not.  If we make them arbitrarily small by
letting $e$ approach zero, we see from the discussion following
Eq.~(5.20) that $m^2$ does not remain in region (1), the region where
spontaneous symmetry breaking occurs; instead, it moves into region
(2) or (3), and the symmetry of the theory remains unbroken.

\vfill \eject

\centerline{References}

\bigskip

\parindent=0pt
\def\bk{\hfill\break\phantom{1.~}}

1. J.~Goldstone, Nuovo Cimento {\bf 19}, 154 (1961); \bk
   Y.~Nambu and G.~Jona-Lasinio, Phys. Rev. {\bf 122}, 345 (1961);\bk
   J.~Goldstone, A.~Salam, and S.~Weinberg, Phys. Rev. {\bf 127},
   965 (1962).

\smallskip
\smallskip

2. F.~Englert and R.~Brout, Phys. Rev. Letters {\bf 13}, 321 (1964);
\bk   P.~Higgs, Phys. Letters {\bf 12}, 132 (1964); 
\bk   G.~S.~Guralnik, C.~R.~Hagen, and T.~W.~B.~Kibble, Phys.~Rev.~Letters~{\bf 13},
 585 (1964);
\bk   P.~Higgs, Phys. Rev. {\bf 145}, 1156 (1966); 
\bk   T.~W.~B.~Kibble, Phys. Rev. {\bf 155}, 1554 (1967);
\bk   G.~'t~Hooft, Nucl. Phys. {\bf B33}, 173 (1971); {\bf B35}, 167 (1971);
\bk   B.~W.~Lee, Phys. Rev. D {\bf 5}, 823, (1972)
\bk   B.~W.~Lee and J.~Zinn-Justin, Phys.~Rev.~D {\bf 5}, 3121 (1972);
{\bf 5}, 3137 (1972); {\bf 5}, 3155 (1972).
\bk   S.~Weinberg, Phys. Rev. D {\bf 7}, 1068 (1973).

\smallskip
\smallskip

3.  Much of this work appear in S.~Coleman and E.~Weinberg, 
Phys. Rev. D {\bf 7}, 1888 (1973).

\smallskip
\smallskip

4. J.~Schwinger, Proc. Natl. Acad. U.S. {\bf 37}, 452 (1951); 
{\bf 37}, 455 (1951); \bk
G.~Jona-Lasinio, Nuovo Cimento {\bf 34}, 1790 (1964). \bk
Functional methods were also used by Goldstone, Salam, and Weinberg
(Ref. 2).

\smallskip
\smallskip

5. Y.~Nambu, Phys. Letters {\bf 26B}, 626 (1968); \bk
   S.~Coleman, J.~Wess, and B.~Zumino, Phys. Rev. {\bf 177}, 2238 (1969).

\smallskip
\smallskip

6. We use the conventions of J.~Bjorken and S.~Drell, {\it Relativistic Quantum 
Fields} (Mc-Graw-Hill, New York, 1965).

\smallskip
\smallskip

7. L.~D.~Faddeev and V.~N.~Popov, Phys. Letters {\bf 25B}, 29 (1967).

\smallskip
\smallskip

8. S.~Weinberg, Phys. Rev. Letters {\bf 19}, 1264 (1967); \bk
A. Salam, in {\it Elementary Particle Theory: Relativistic Groups and 
Analyticity} (Nobel Symposium No. 8), edited by N. Svartholm (Wiley,
New York, 1969).

\smallskip
\smallskip

9. G.~Feldman and P.~T.~Mathews, Phys. Rev. {\bf 130}, 1633 (1963).

\smallskip
\smallskip

10.  S.~Weinberg, Phys. Rev. Letters {\bf 29}, 1698 (1972).

\smallskip
\smallskip

\def\bk{\hfill\break\phantom{11.~}}

11. M.~Gell-Mann and F.~Low, Phys. Rev. {\bf 95}, 1300 (1954); \bk
  N.~N.~Bogoliubov and D.~V.~Shirkov, {\it Introduction to the Theory of 
Quantized Fields} (Interscience, New York, 1959); \bk
C.~G.~Callan, Phys. Rev. D {\bf 2}, 1541 (1970); \bk
K.~Symanzik, Commun. Math. Phys. {\bf 18}, 227 (1970).

\smallskip
\smallskip

12. See the contribution of S.~Coleman to Proceedings of the 1971
International Summer School ``Ettore Majorana'' (Academic, New York, to
be published).

\smallskip
\smallskip

13. G.~'t~Hooft and M.~Veltman, Nucl. Phys. {\bf B44}, 189 (1972).

\smallskip
\smallskip

14.  H.~Georgi and S.~Glashow, Phys. Rev. D {\bf 6}, 2977 (1972).

\vfill\eject

\centerline{Figures}

\begin{figure}[h]
\begin{center}
\scalebox{0.5}[0.5]{\includegraphics{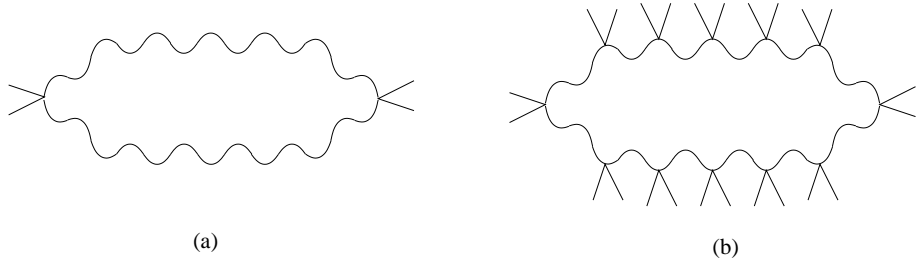}}
\par
\vskip-2.0cm{}
\end{center}
\begin{quote}
\caption{\small Two diagrams which contribute to the effective
potential in scalar electrodynamics.  The wiggly lines represent
photons, while the straight lines represent scalar mesons. }
\end{quote}
\end{figure}

\begin{figure}[h]
\begin{center}
\scalebox{0.5}[0.5]{\includegraphics{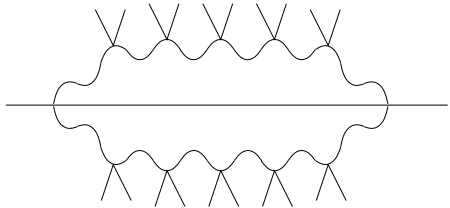}}
\par
\vskip-2.0cm{}
\end{center}
\begin{quote}
\caption{\small  A two-loop diagram which contributes to the 
effective potential in scalar electrodynamics.}
\end{quote}
\end{figure}

\clearpage
\phantom{here}

\begin{figure}[t]
\begin{center}
\scalebox{0.1}[0.1]{\includegraphics{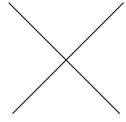}}
\par
\vskip-2.0cm{}
\end{center}
\begin{quote}
\caption{\small The only zero-loop diagram in the $\lambda\phi^4$ model.}
\end{quote}
\end{figure}


\begin{figure}[ht]
\begin{center}
\scalebox{0.7}[0.7]{\includegraphics{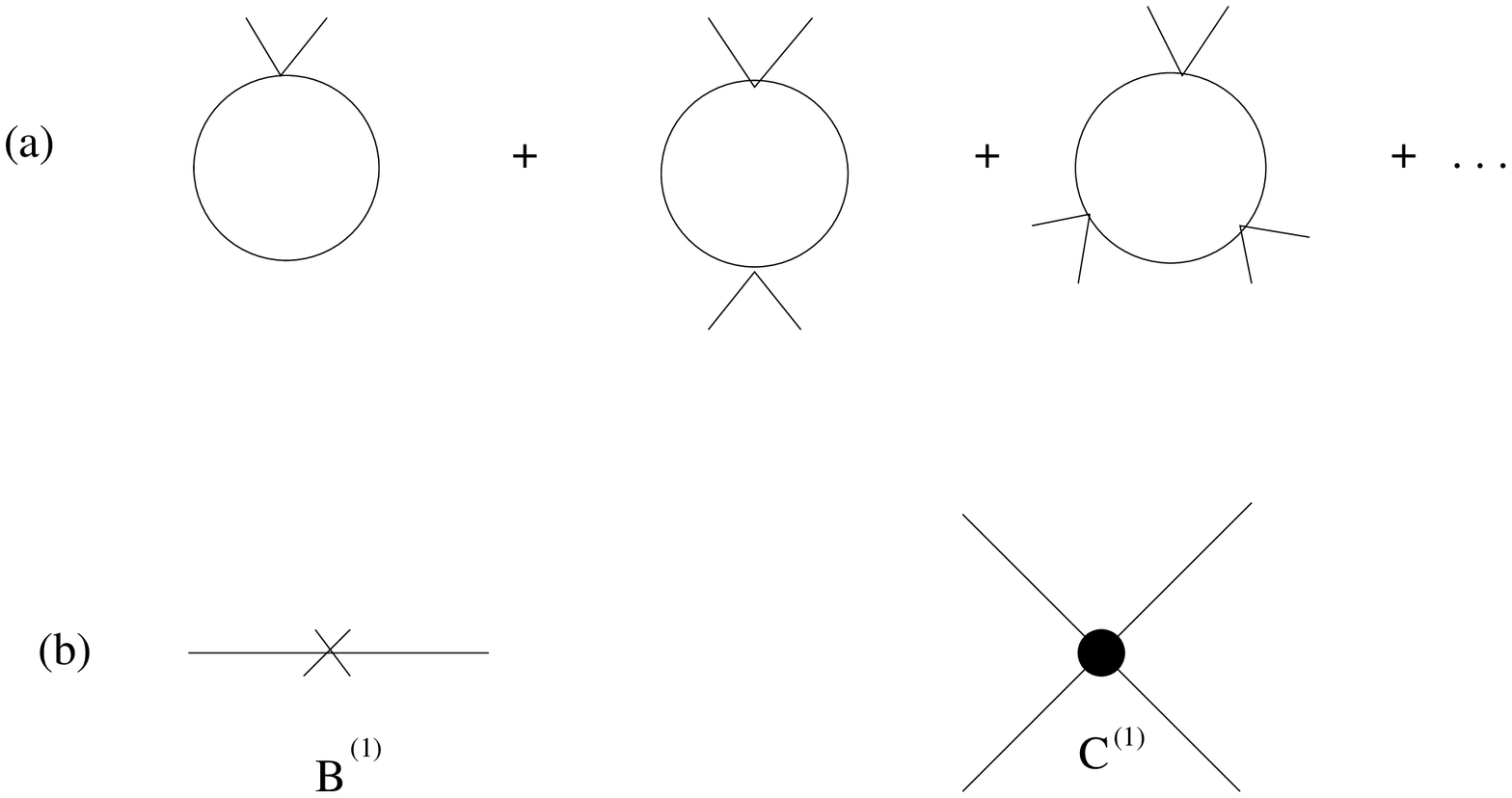}}
\par
\vskip-12.0cm{}
\end{center}
\begin{quote}
\caption{\small The one-loop diagrams which contribute to the
effective potential in the $\lambda \phi^4$ model.}
\end{quote}
\end{figure}

\clearpage

\begin{figure}[t]
\begin{center}
\scalebox{0.7}[0.5]{\includegraphics{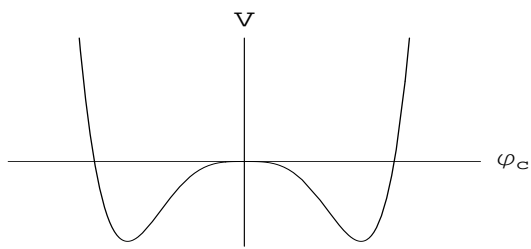}}
\par
\vskip-2.0cm{}
\end{center}
\begin{quote}
\caption{\small The behavior of the effective potential in many models.}
\end{quote}
\end{figure}

\begin{figure}[b]
\begin{center}
\scalebox{0.6}[0.7]{\includegraphics{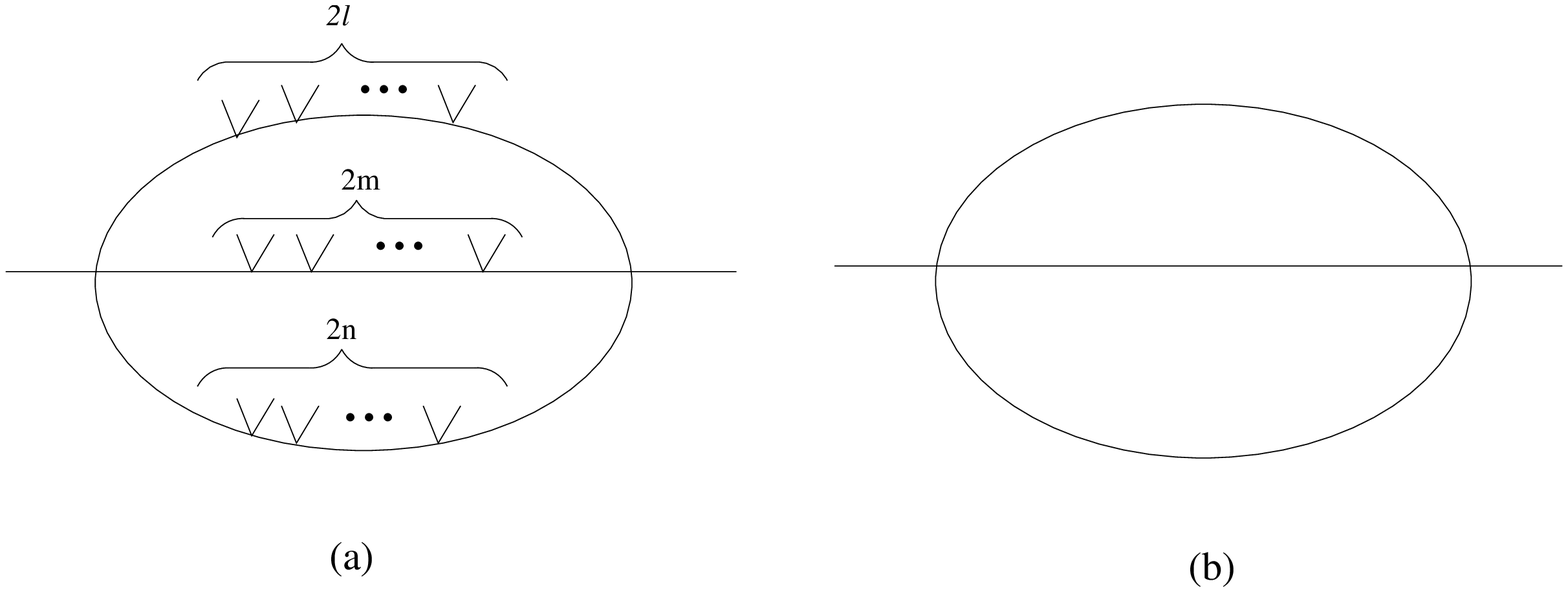}}
\par
\vskip-2.0cm{}
\end{center}
\begin{quote}
\caption{\small a) A two-loop graph occurring in the $\lambda \phi^4$ model.
       \hfill \break 
{}\phantom{Figure 6. }  b) The corresponding prototype graph.}
\end{quote}
\end{figure}

\clearpage

\begin{figure}[ht]
\begin{center}
\scalebox{1.0}[1.0]{\includegraphics{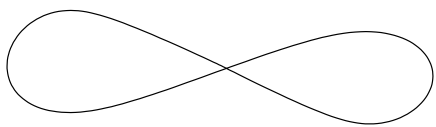}}
\par
\vskip-2.0cm{}
\end{center}
\begin{quote}
\caption{\small A prototype graph which contributes to the
effective potential in the $\lambda \phi^4$ model. }
\end{quote}
\end{figure}

\begin{figure}[ht]
\begin{center}
\scalebox{0.7}[0.7]{\includegraphics{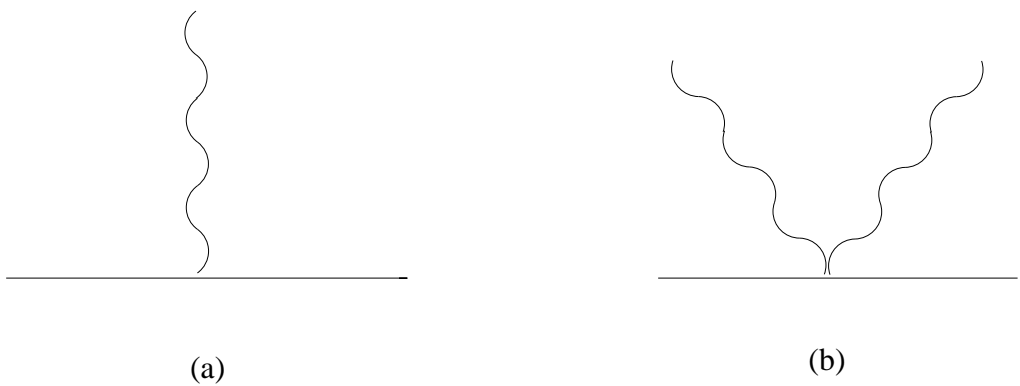}}
\par
\vskip-2.0cm{}
\end{center}
\begin{quote}
\caption{\small The two types of scalar-vector vertices in gauge
theories.  The wiggly lines represent vector mesons, the straight
lines represent scalar mesons.}
\end{quote}
\end{figure}

\clearpage

\begin{figure}[ht]
\begin{center}
\scalebox{0.7}[0.7]{\includegraphics{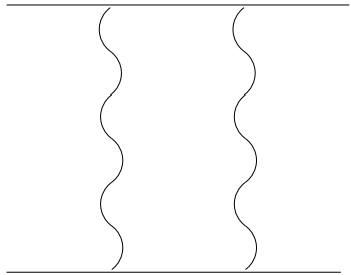}}
\par
\vskip-2.0cm{}
\end{center}
\begin{quote}
\caption{\small A diagram which does not contribute to the effective
potential in Landau gauge.  }
\end{quote}
\end{figure}

\begin{figure}[ht]
\begin{center}
\scalebox{0.9}[0.7]{\includegraphics{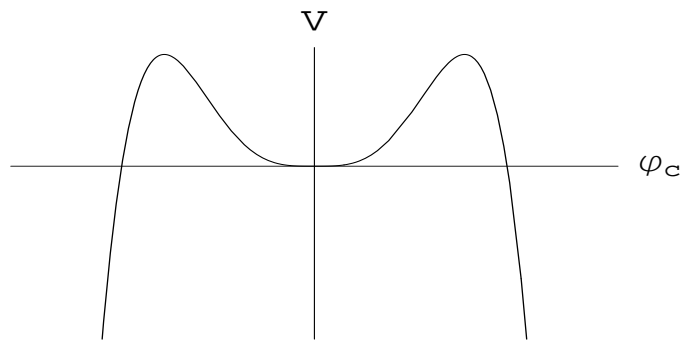}}
\par
\vskip-2.0cm{}
\end{center}
\begin{quote}
\caption{\small The behavior of the effective potential in the 
Yukawa model when $g^4 > \lambda^2/16$.   }
\end{quote}
\end{figure}

\begin{figure}[ht]
\begin{center}
\scalebox{1.0}[1.0]{\includegraphics{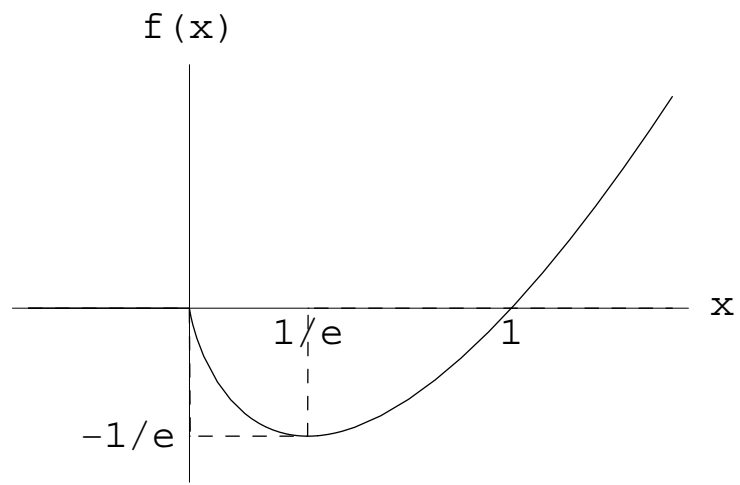}}
\par
\vskip-2.0cm{}
\end{center}
\begin{quote}
\caption{\small The function $f(x) = x \log(x)$.  }
\end{quote}
\end{figure}

\clearpage

\begin{figure}[t]
\begin{center}
\scalebox{0.9}[0.7]{\includegraphics{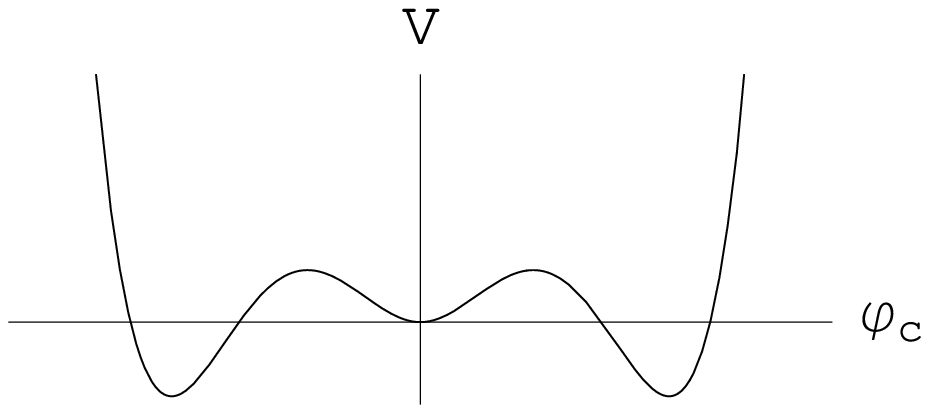}}
\par
\end{center}
\begin{center}
{(a) $0 < m^2 < {3e^4 \mu^2 \over 32\pi^2}
e^{-1/2}$  }
\end{center}

\begin{center}
\scalebox{0.9}[0.7]{\includegraphics{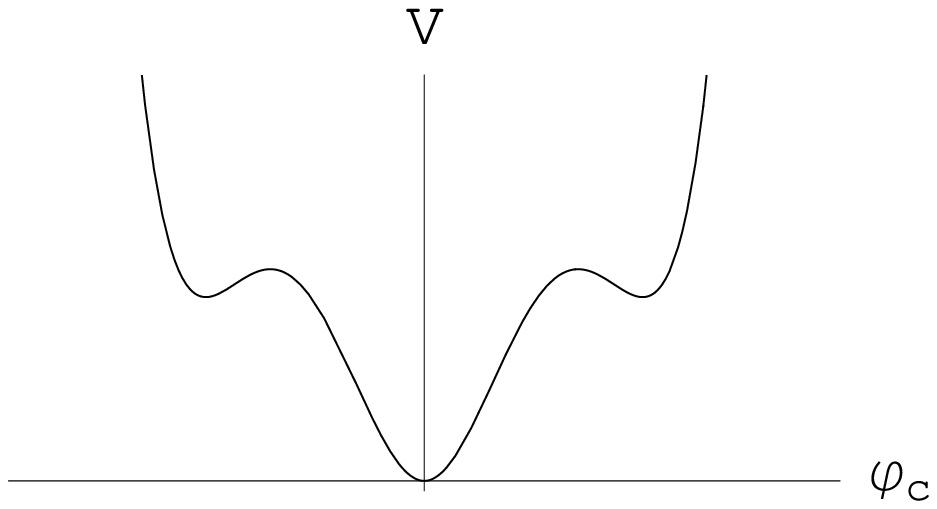}}
\par
\end{center}
\begin{center}
{ (b) ${3e^4 \mu^2 \over 32\pi^2}e^{-1/2}
 <m^2 < {3e^4 \mu^2 \over 32\pi^2}e^{-1}$}
\end{center}

\begin{center}
\scalebox{0.9}[0.7]{\includegraphics{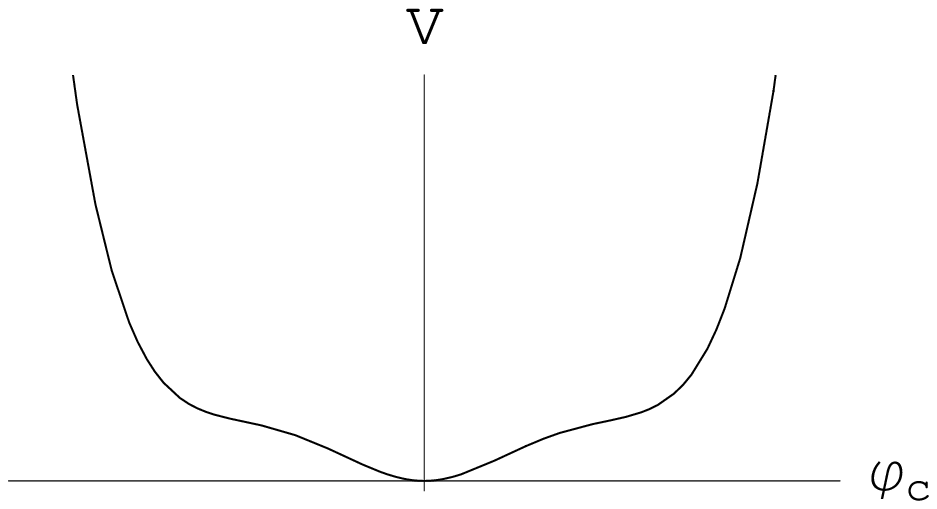}}
\par
\end{center}
\begin{center}
{ (c) $m^2 > {3e^4 \mu^2 \over 32\pi^2}e^{-1}$}
\end{center}
\begin{quote}
\caption{\small The behavior of the effective potential in massive
scalar electrodynamics. }

\end{quote}
\end{figure}

\clearpage

\begin{figure}[ht]
\begin{center}
\scalebox{1.0}[1.0]{\includegraphics{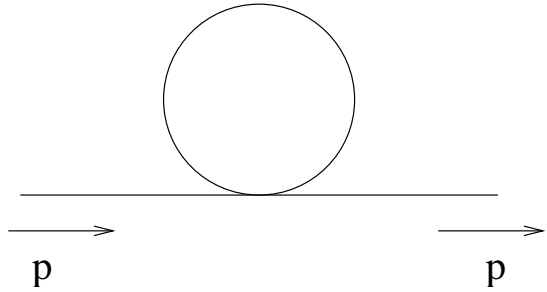}}
\par
\vskip-2.0cm{}
\end{center}
\begin{quote}
\caption{\small The only diagram contributing to $Z(\pc)$ in the 
$\lambda \phi^4$ model in the one-loop approximation. }
\end{quote}
\end{figure}

\begin{figure}[hb]
\begin{center}
\scalebox{0.5}[0.7]{\includegraphics{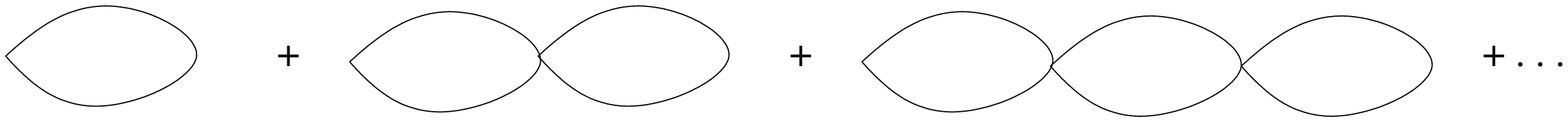}}
\par
\vskip-2.0cm{}
\end{center}
\begin{quote}
\caption{\small The most important prototype graphs for the 
calculation of the effective potential in the scalar SO($n$) model.  }
\end{quote}
\end{figure}

\begin{figure}[t]
\begin{center}
\scalebox{0.7}[0.7]{\includegraphics{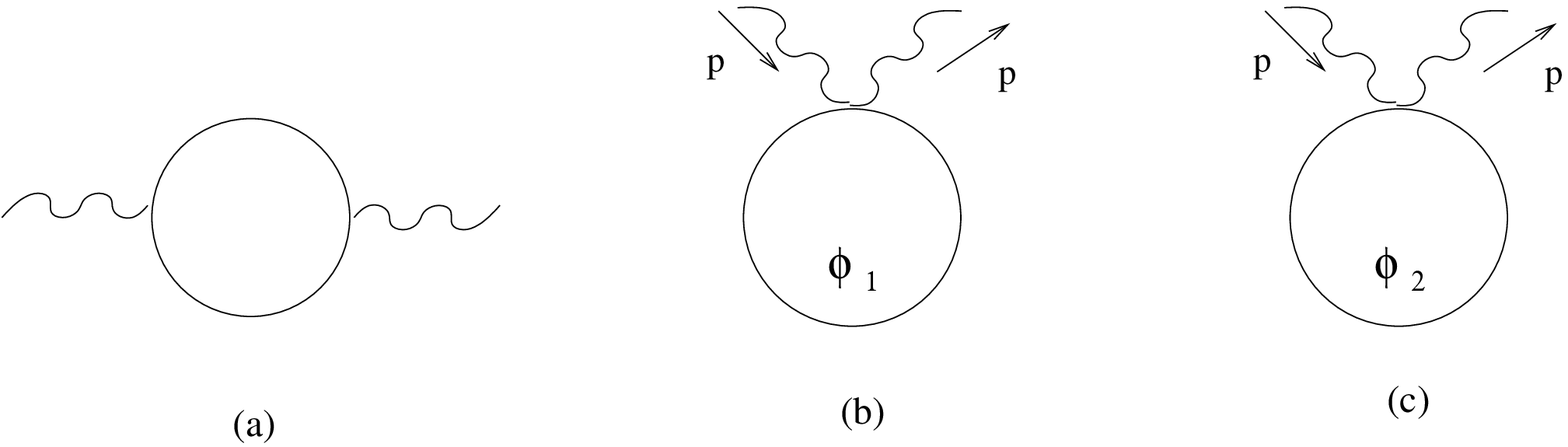}}
\par
\vskip-2.0cm{}
\end{center}
\begin{quote}
\caption{\small The graphs which contribute to $H(\pc)$ in scalar
electrodynamics in the one-loop approximation.  }
\end{quote}
\end{figure}


\begin{figure}[b]
\begin{center}
\scalebox{0.65}[0.65]{\includegraphics{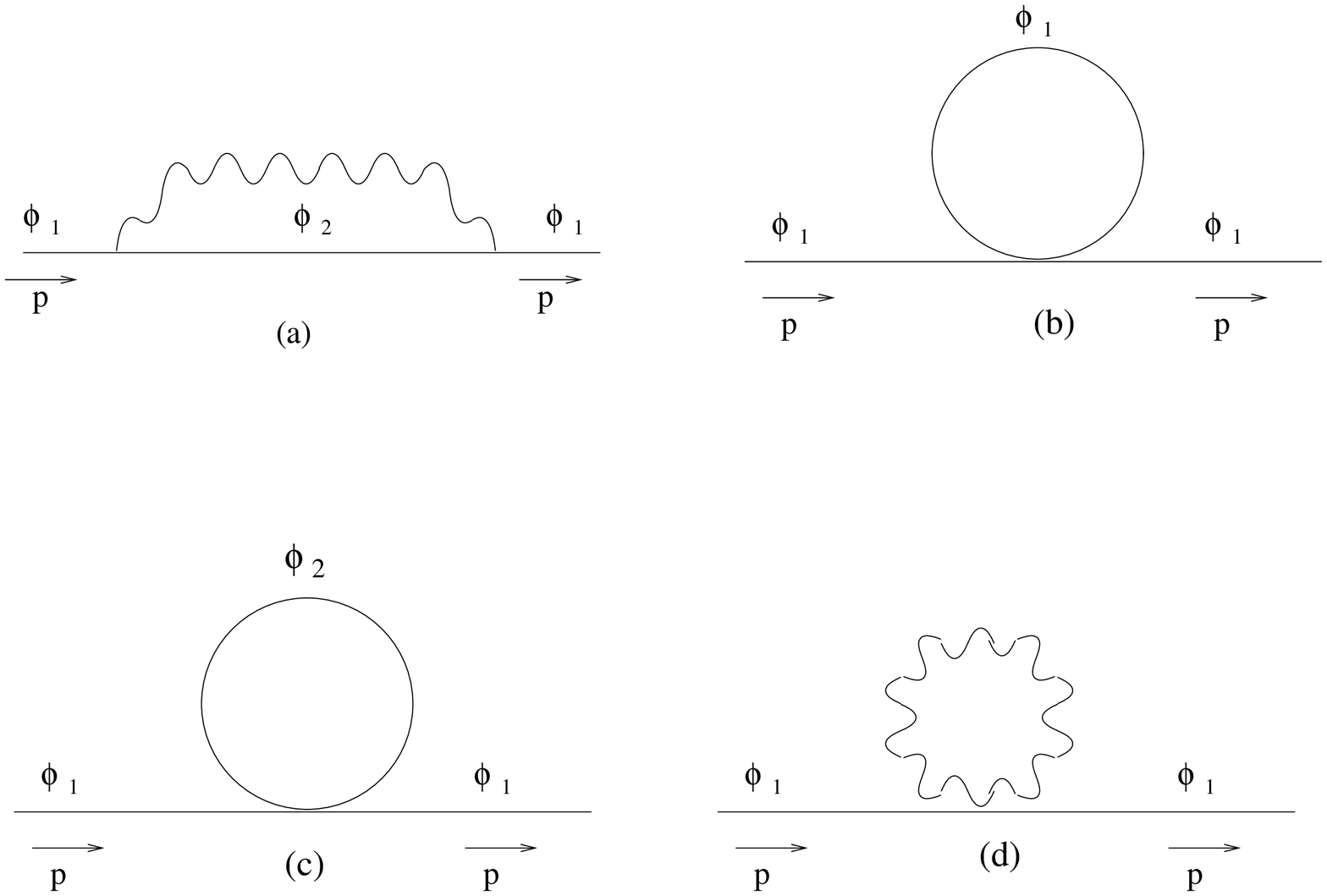}}
\par
\vskip-2.0cm{}
\end{center}
\begin{quote}
\caption{\small The graphs which contribute to $Z(\pc)$ in scalar
electrodynamics in the one-loop approximation. }
\end{quote}
\end{figure}


\begin{figure}[ht]
\begin{center}
\hskip-1cm\scalebox{0.65}[0.65]{\includegraphics{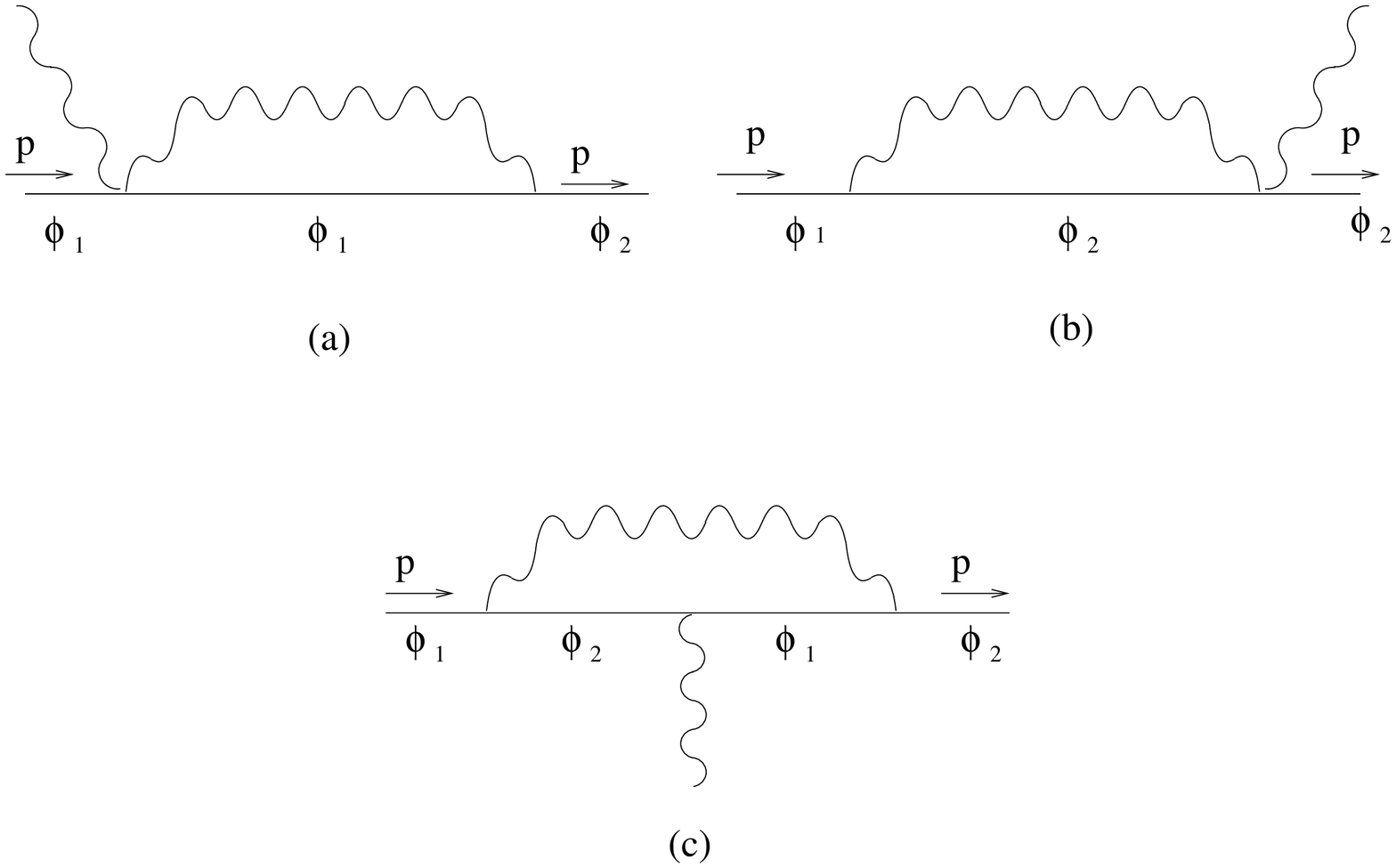}}
\par
\vskip-2.0cm{}
\end{center}
\begin{quote}
\caption{\small  The graphs which contribute to $F(\pc)$ in scalar
electrodynamics in the one-loop approximation.  }
\end{quote}
\end{figure}

\clearpage

\begin{figure}[t]
\begin{center}
\scalebox{0.7}[0.7]{\includegraphics{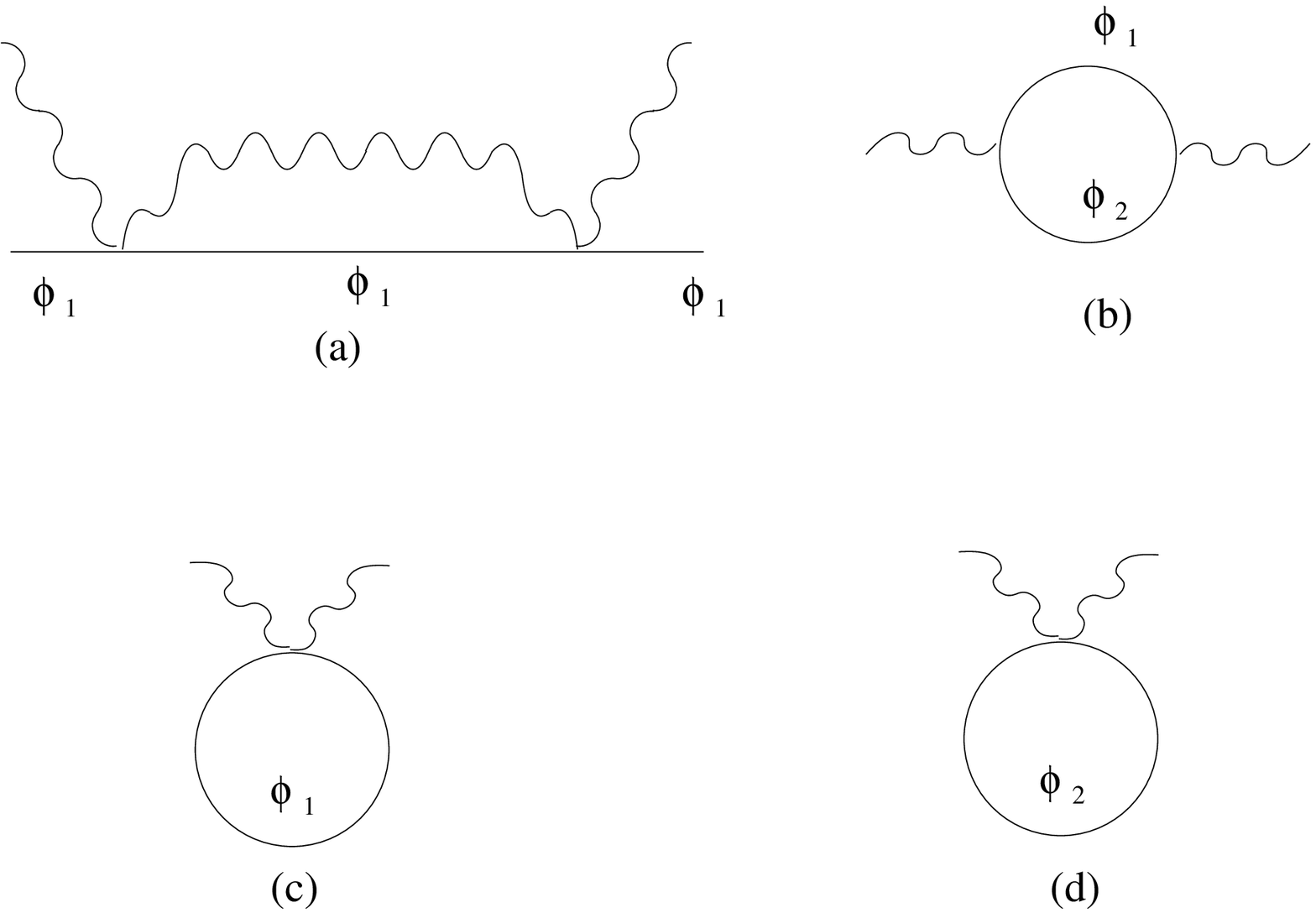}}
\par
\vskip-2.0cm{}
\end{center}
\begin{quote}
\caption{\small  The graphs which contribute to $G(\pc)$ in scalar
electrodynamics in the one-loop approximation.  }
\end{quote}
\end{figure}

\begin{figure}[t]
\begin{center}
\scalebox{0.7}[0.7]{\includegraphics{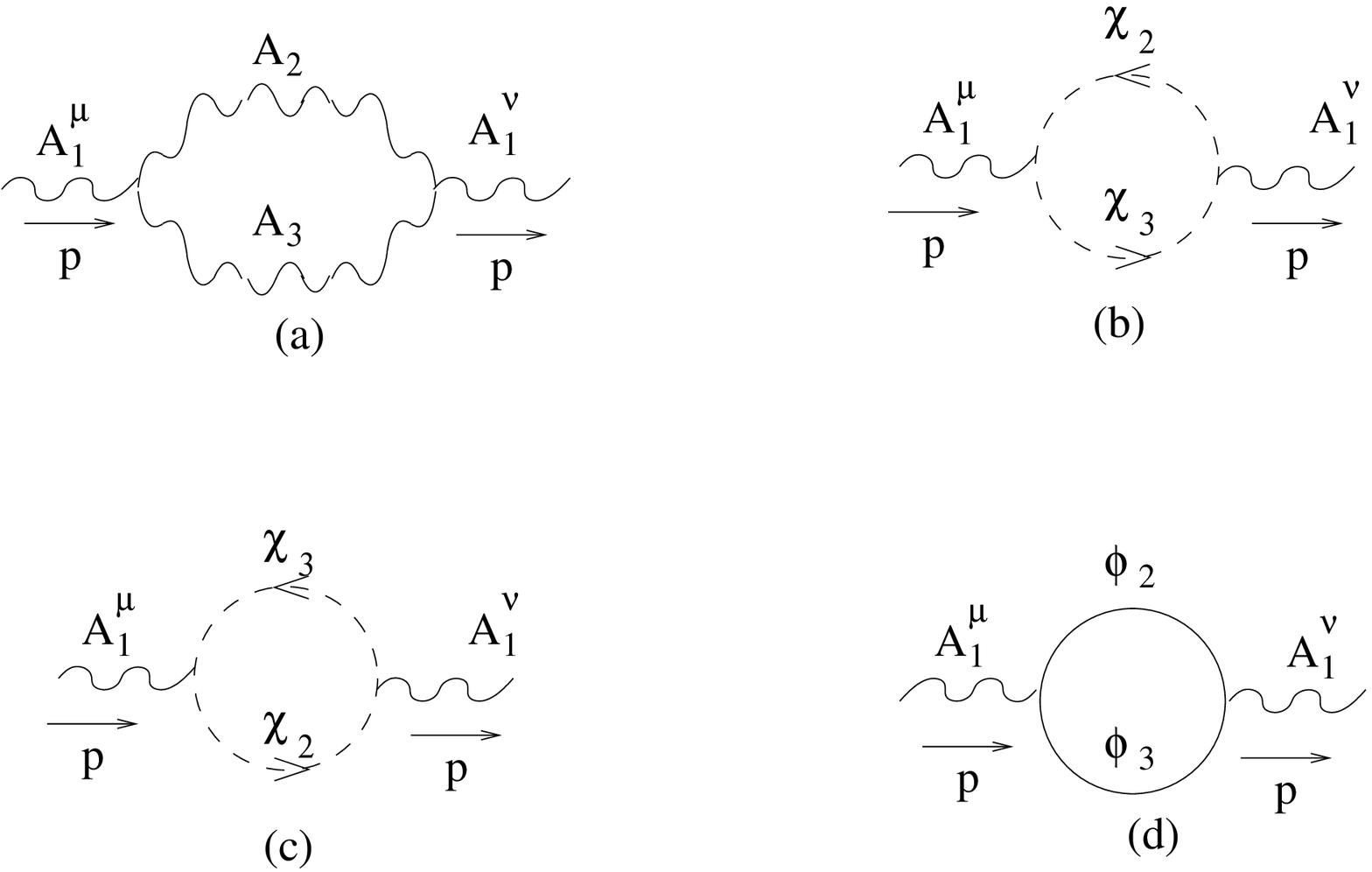}}
\par
\vskip-2.0cm{}
\end{center}
\begin{quote}
\caption{\small The one-loop graphs which give a non-vanishing 
contribution to $H(\pc)$ in the massless Yang-Mills model.  }
\end{quote}
\end{figure}

\begin{figure}[tt]
\begin{center}
\scalebox{0.69}[0.7]{\includegraphics{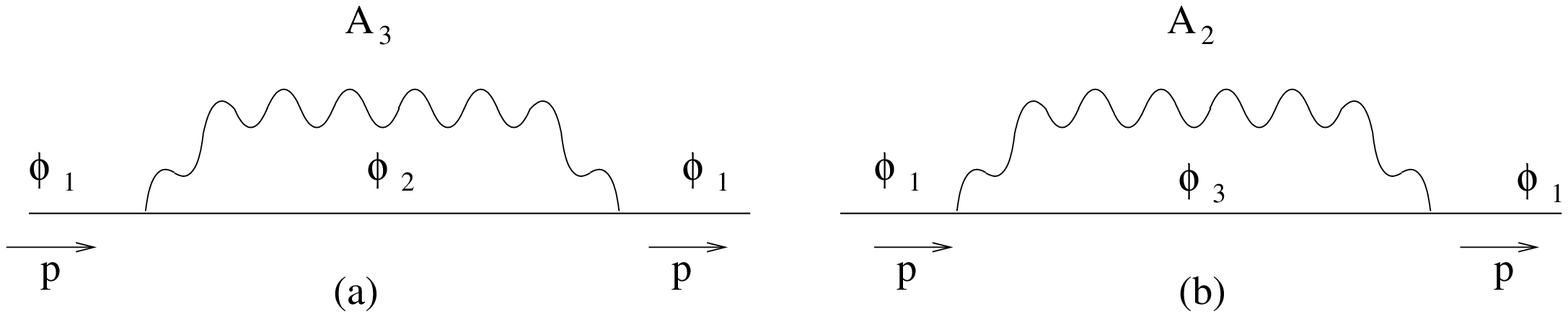}}
\par
\vskip-2.0cm{}
\end{center}
\begin{quote}
\caption{\small The one-loop graphs which give a non-vanishing 
contribution to $Z(\pc)$ in the massless Yang-Mills model.  }
\end{quote}
\end{figure}

\begin{figure}[t]
\begin{center}
\scalebox{0.64}[0.7]{\includegraphics{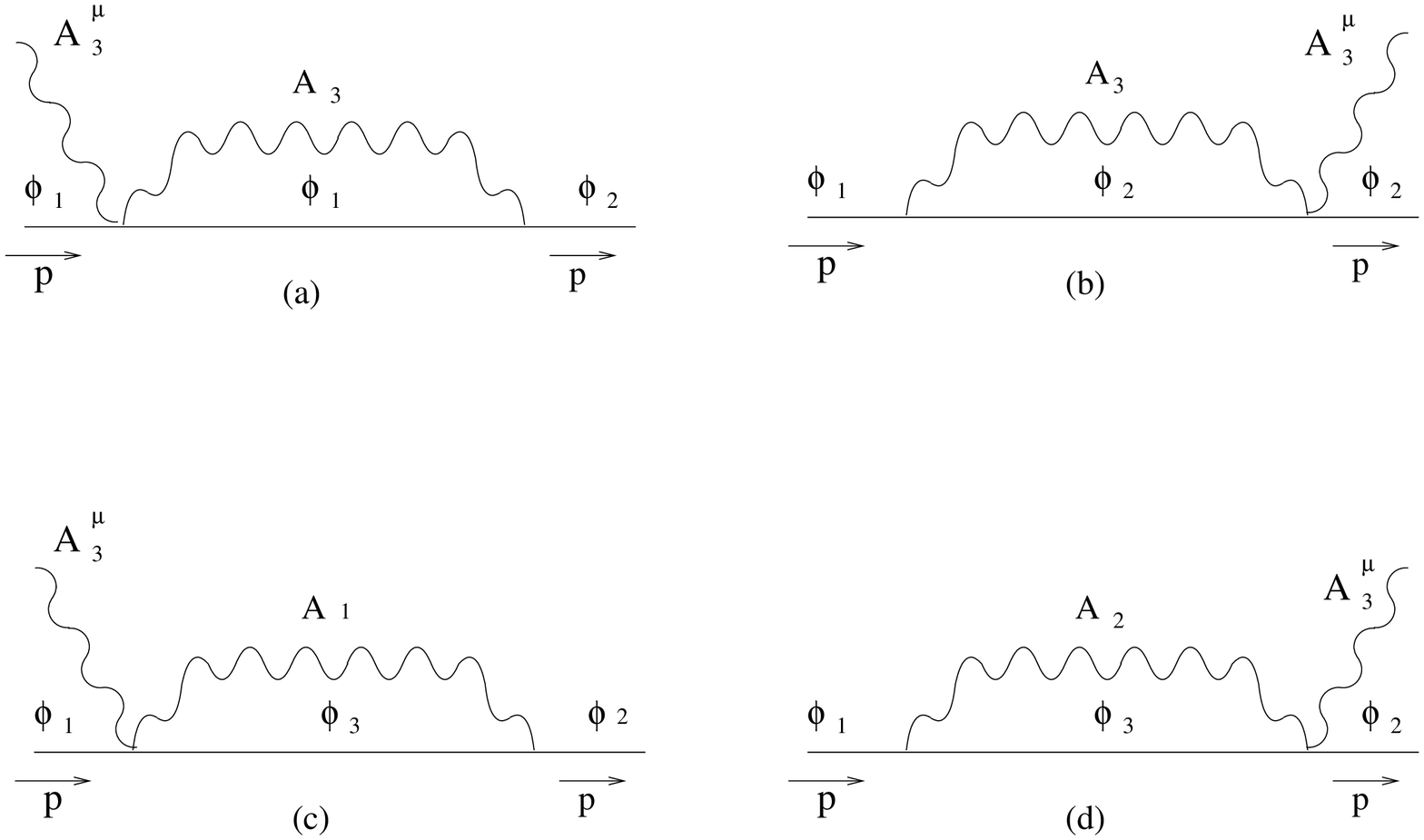}}
\par
\vskip-2.0cm{}
\end{center}
\begin{quote}
\caption{\small The one-loop graphs which give a non-vanishing 
contribution to $F(\pc)$ in the massless Yang-Mills model.  }
\end{quote}
\end{figure}

\end{document}